\documentclass{aa}
\usepackage{graphicx}
\usepackage{txfonts}
\usepackage{placeins}
\usepackage{hyperref}
\usepackage{siunitx}
\usepackage{multirow}
\usepackage{afterpage}
\hypersetup{
 colorlinks   = true, 
 urlcolor     = blue, 
 linkcolor    = blue, 
 citecolor    = blue  
 }

\usepackage[normalem]{ulem}
\newcommand{\dg}{$^\circ$}

\newcommand{\rs}{R$_{\odot}$}

\begin{document}

\title{Five new eclipsing binaries with low-mass companions}

\author{ J. Lipt\'ak~\inst{1,2}
\and M. Skarka~\inst{1,4}
\and E. Guenther~\inst{3}
\and P. Chaturvedi~\inst{3}
\and M. V\'itkov\'a~\inst{1,4}
\and R. Karjalainen~\inst{1}
\and J.~Šubjak~\inst{1,9}
\and A.~Hatzes~\inst{3}
\and A.~Bieryla~\inst{9}
\and D.~Gandolfi~\inst{10}
\and S. H.\ Albrecht\inst{13}
\and P.~G.~Beck~\inst{5,6}
\and H. J.~Deeg~\inst{5,6}
\and M.~E.~Everett~\inst{8}
\and J.~Higuera~\inst{8}
\and D.~Jones~\inst{5,6,7}
\and S.~Mathur~\inst{5,6}
\and Y.~G.~Patel~\inst{8}
\and C.~M.~Persson~\inst{11}
\and S.~Redfield~\inst{12}
\and P.~Kab\'ath~\inst{1}
}

\institute{ Astronomical Institute, Academy of Sciences, Fri\v{c}ova~298, 251~65~Ond\v{r}ejov, Czech Republic  
\and Astronomical Institute, Faculty of Mathematics and Physics, Charles University, V~Hole\v{s}ovi\v{c}k\'ach~2, 180~00~Praha~8, \\Czech Republic
\and Thüringer Landessternwarte Tautenburg, Sternwarte 5, 07778 Tautenburg, Germany  
\and Department of Theoretical Physics and Astrophysics, Faculty of Science,
Masaryk University, Kotl\'a\v{r}sk\'a 267/2, 611~37 Brno, \\Czech Republic
\and Instituto de Astrof\'isica de Canarias, E-38205 La Laguna, Tenerife, Spain
\and Departamento de Astrof\'isica, Universidad de La Laguna, E-38206 La Laguna, Tenerife, Spain
\and Nordic Optical Telescope, Rambla Jos\'e Ana Fern\'andez P\'erez 7, 38711, Bre\~na Baja, Spain
\and NSF’s National Optical-Infrared Astronomy Research Laboratory, 950 N. Cherry Ave., Tucson, AZ 85719, USA
\and Center for Astrophysics ${\rm \mid}$ Harvard {\rm \&}
Smithsonian, 60 Garden Street, Cambridge, MA 02138, USA
\and Dipartimento di Fisica, Universitá di Torino, via P. Giuria 1, 10125 Torino, Italy
\and Department of Space, Earth and Environment, Chalmers University of Technology, Onsala Space Observatory, SE-439 92 Onsala, Sweden
\and Astronomy Department and Van Vleck Observatory, Wesleyan University, Middletown,
CT 06459, USA
\and Stellar Astrophysics Centre, Department of Physics and 
Astronomy, Aarhus University, Ny Munkegade 120, DK-8000 Aarhus C, Denmark}
\date{Received \today}

\abstract{Precise space-based photometry from the Transiting Exoplanet Survey Satellite results in a huge number of exoplanetary candidates. However, the masses of these objects are unknown and must be determined by ground-based spectroscopic follow-up observations, frequently revealing the companions to be low-mass stars rather than exoplanets. We present the first orbital and stellar parameter solutions for five such eclipsing binary-star systems using radial-velocity follow-up measurements together with spectral-energy-distribution solutions. TOI-416 and TOI-1143 are totally eclipsing F+M star systems with well-determined secondary masses, radii, and temperatures. TOI-416 is a circular system with an F6 primary and a secondary with a mass of $M_2=\SI{0.131(8)}{M_\odot}$. TOI-1143 consists of an F6 primary with an $M_2={0.142(3)}{M_\odot}$ secondary on an eccentric orbit with a third companion. With respect to the other systems, TOI-1153 shows ellipsoidal variations, TOI-1615 contains a pulsating primary, and TOI-1788 has a spotted primary, while all have moderate mass ratios of 0.2-0.4. However, these systems are in a grazing configuration, which limits their full description. The parameters of TOI-416B and TOI-1143B are suitable for the calibration of the radius-mass relation for dwarf stars.
 }


\maketitle

\section{Introduction}
Dwarf stars of spectral type M are the most common in the Galaxy. With masses of $M<\SI{0.6}{M_\odot}$ and above the hydrogen burning limit of $M=\SI{0.075}{M_\odot}$ \citep{2023A&A...671A.119C}, they comprise up to $\approx 40\%$ of the entire stellar population assuming initial mass-function by \cite{2003PASP..115..763C}. However, our knowledge of their fundamental parameters is still limited. To estimate the masses of single M dwarfs, one needs to use stellar structure models. The only quantities directly related to the observations are the effective temperature and stellar radius derived from the spectral energy distribution (SED) of dwarfs with measured parallaxes and metalicity, and surface gravity derived from spectroscopy. However, the precision of the surface gravities determined in this way is on the order of 10-15\%\ at best, which is not low enough to constrain the mass well.

The physics of stellar modeling of low-mass stars is challenging. These stars are dominated by convection; stars with masses $\lesssim \SI{0.35}{M_\odot,}$ are fully convective according to \cite{1997A&A...327.1039C}. This, together with the rotation of the stars, generates magnetic fields. Further complication comes from the boundary conditions - cold atmospheres contain molecular species with many transitions contributing significantly to the opacity. For the calibration of these models, we need observations with sufficiently small uncertainties on the masses and radii. The only source of mass estimates that are independent of stellar structure models are binary stars, either astrometric or eclipsing, where we can obtain the secondary mass from radial velocities. In the case of a purely spectroscopic binary (i.e.,\ not eclipsing), we can only determine $M_2 \sin i$, where $i$ is the inclination of the binary orbit. 

Before 2010, only about a dozen M dwarfs with masses and radii with levels of precisions better than $10\%$ were known, usually M+M systems close to Earth such as CU Cnc \citep{2003A&A...398..239R}. Subsequently, space-based exoplanetary missions came to the rescue. The eclipse depth of an M dwarf star orbiting an early-type star is quite similar to a brown dwarf or a hot Jupiter orbiting a G, K, or M star. Thus, the mission pipelines were automatically identifying these systems. Further spectroscopic follow-up is needed to discern the cases based on radial-velocity amplitudes. This led to an increase in interest. The number of these binary systems is steadily growing, with the current count being just below 100 thanks to works such as the EBLM (eclipsing binary–low mass) project \citep{2013A&A...549A..18T}, \cite{2023Univ....9..498M} using WASP and Kepler candidates, \cite{2018AJ....156...27C} using different photometry sources, and \cite{2018AcA....68..449A} studying contact binary systems and others. These new data started to show systematic deviations from the models of \cite{2015A&A...577A..42B} or MIST \citep{2016ApJ...823..102C}. For a given mass, the observational radii are around $5\%$ bigger, and temperatures are around $\SI{100}{K}$ colder than expected, as was shown in works of \citet{2018MNRAS.481.1083P}, \citet{2023MNRAS.521.6305D}, and \citet{2023ApJ...951...90W}. Also, there is a larger scatter in the radius-mass relation than expected, suggesting that magnetic fields, rotation, and detailed chemical composition may play a role that is not well described by the models. 

Most stars in the Galaxy form with companions. About half of all main-sequence stars of spectral classes F and G have smaller companions, with about half of those being close binaries (with $a<\SI{10}{au}$) as is shown in the review on stellar multiplicity by \citet{2023ASPC..534..275O}. Concerning the orbital eccentricities of the close binaries, the internal structure of stars is important. Stars below the Kraft break ($M\approx \SI{1.3}{M_\odot}$) \citep{1967ApJ...150..551K} with convective zones in their interior are expected to be circularized below $P=\SI{8}{days}$ in comparison with hotter, radiative-envelope stars that tend to have circular orbits only below $P=\SI{2}{days}$ \citep{2023ASPC..534..275O}. Among solar type stars, the distribution of mass ratios is quite uniform, except for the overabundance of twin systems and the low-mass cut-off at brown dwarf masses. This underabundance of close brown-dwarf companions was described by \cite{1988ApJ...331..902C} and is commonly called the brown-dwarf desert. The most probable cause is a difference in the mechanisms that drive the formation of close stellar companions (disk fragmentation) and planets (core accretion) that likely overlap in the brown-dwarf mass range. More recently, the imaging survey by \citet{2023MNRAS.519..778D} extended the brown-dwarf desert to an absence of low-mass stellar companions to late B- and A-type stars (albeit just in wide binaries), suggesting that the processes are mostly invariant of primary mass and driven by mass ratio ($q$) instead, with a truncation of companion distribution at $q=0.075$.

Here, we present the characterization of five systems identified as exoplanetary candidates via the Transiting Exoplanet Survey Satellite (TESS) mission that turned out to be eclipsing binaries with low-mass stellar companions. TOI-416 and TOI-1143 are well-described systems comprising F-type primaries and late M-dwarf secondaries. These are TOI-1615 -- an M dwarf secondary at the convective-radiative core transition orbiting around a $\delta$ Scuti variable, and TOI-1153 and TOI-1788, which are are systems with secondaries of late-K and mid-M spectral type for which the description is somewhat limited by the grazing nature of the eclipses.

\section{Observations}
\label{sec:obs}
Our targets were identified as exoplanetary candidates in TESS space-based photometry between January 2019 and March 2020. We contributed to the ground-based spectroscopic follow-up observations revealing the systems to be stellar binaries. Further, we use these measurements to constrain the physical properties of the stellar components of the systems and their orbital parameters. All the radial-velocity measurements used are listed in Table \ref{Tab:RVs} in the appendix. 

\subsection{TESS photometry}
\label{sec:tess}

The TESS  \citep{2015JATIS...1a4003R} is a photometric space-based observatory operated by NASA, orbiting the Earth on a highly eccentric orbit with an orbital period of half a month. The satellite covers the sky in sectors of 24 $\times$ 96 degrees using wide-field cameras with 105 mm aperture, a focal ratio of f/1.4, and a spatial resolution of about half an arcminute. Each sector is covered continuously for two consecutive orbits with a brief interruption during the perigee used for data download. The survey was designed for the discovery of exoplanets via the transit method using a detector with a broad band pass around 6000 -- 10\,000~\AA.

\begin{figure*}[]
    \centering
    \includegraphics[width=\linewidth]{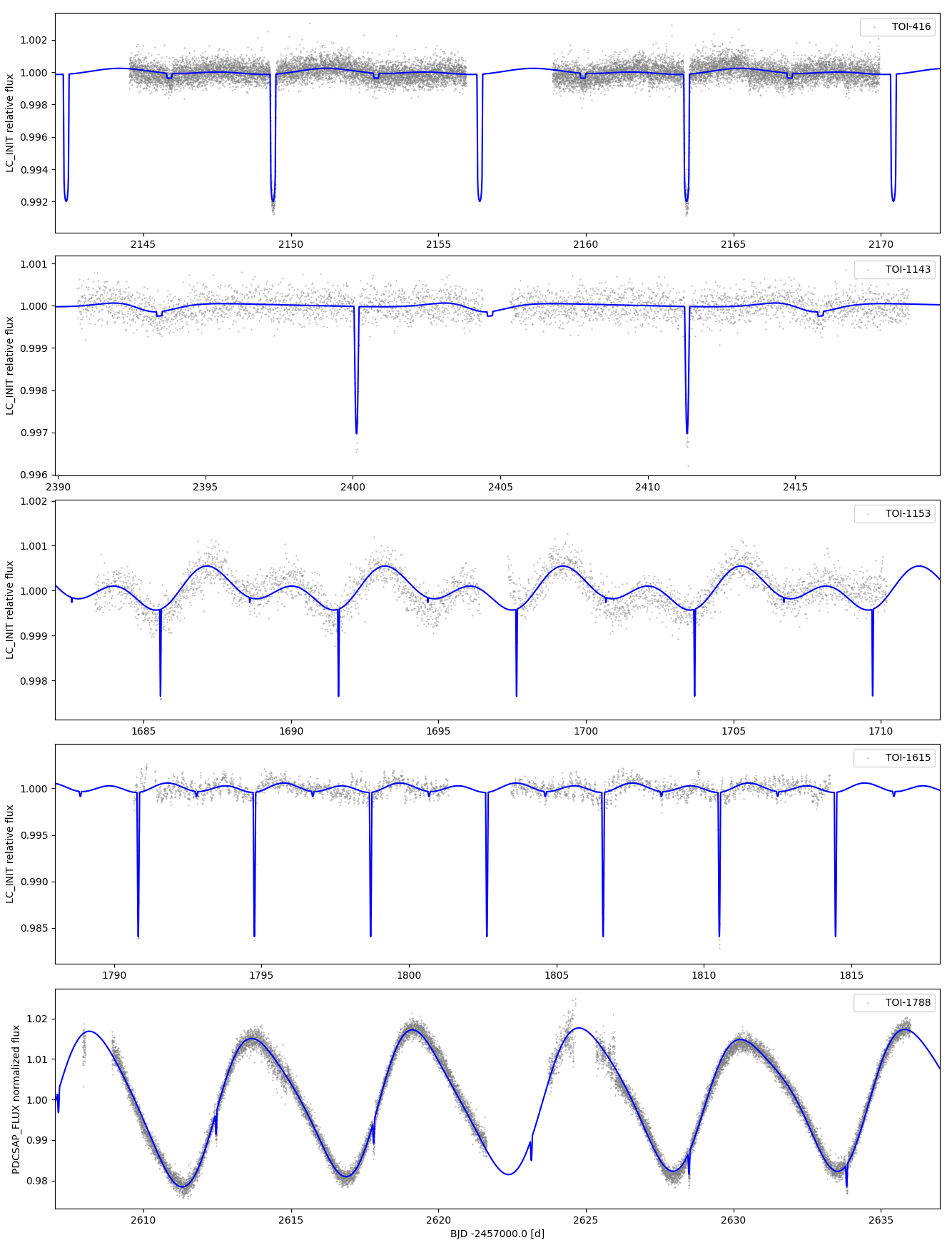}
    \caption{TESS light curves of the systems analyzed here. We only show one example sector per target. From top to bottom, TOI-416 (Sect. 31), TOI-1143 (Sect. 40), TOI-1153 (Sect. 14), TOI-1615 (Sect. 18), and TOI-1788 (Sect. 48). The blue line shows our binary fits with Doppler beaming included; in the case of TOI-1788, we also included model of rotational variability.}
    \label{fig:TESS}
\end{figure*}

We used the MAST~\footnote{Barbara A. Mikulski Archive for Space Telescopes, \url{https://mast.stsci.edu/portal/Mashup/Clients/Mast/Portal.html}} archive to download the available short-cadence photometric time series as of September 2023. For our analysis, we used the LC\_{}INIT time series of relative fluxes from the Data Validation Time Series product. These contain harmonics (of the satellite orbital period) removed, the sector edge de-trended, normalized and stitched light curves from all available sectors as outlined in the TESS Science Data Products Description  Document\footnote{Available via \url{https://archive.stsci.edu/missions-and-data/tess}}. To model light curves it is necessary to add one to these values as the LC\_{}INIT is shifted to an average normalized flux of zero. In cases where the LC\_{}INIT was not yet available or where the light curve was visibly disturbed by the removal of harmonics (in the case of TOI-1788), we used PDCSAP\_{}FLUX fluxes from the relevant Light Curve Files, which we normalized by the mean value to obtain relative fluxes that can be appended to the previous data sets. We show one example sector of the photometric data used for each target in Fig. \ref{fig:TESS}. The number of sectors in which the target was observed, $N_\mathrm{sectors}$; primary eclipses contained within them, $N_\mathrm{prim. ecl.}$; visibility of secondary eclipses; and other features in the light curves can be found in Table~\ref{Tab:LogTESS}.

\begin{table}[h!]
\caption{Available TESS short-cadence photometry used in the analysis.}
\label{Tab:LogTESS}
\centering
\begin{tabular}{c|cccc}
System & $N_\mathrm{sectors}$ & $N_\mathrm{prim. ecl.}$& sec. eclipse&note\\
\hline
TOI-416&  2  & 5  & weak  &   \\
TOI-1143& 4 & 9  & weak  &  $^a$ \\
TOI-1153& 7  & 30  & very week  &   $^b$\\
TOI-1615& 5  & 29  & weak   & $^c$  \\
TOI-1788& 1  & 4  & no  &  $^d$  \\
\end{tabular}

\footnotesize $^a$shows an eccentric orbit, $^b$asymmetric out of eclipse, \\$^c$small, $\delta$-Sct-like variations at around~6~c/d,\\ $^d$dominated by sinusoidal variation at $P=\SI{5.55}{d}$
\end{table} 

We checked the apertures used for light-curve extraction for field star contamination. Usually, we found no more than two sources with magnitude contrast $\Delta m <7$. As an exception, TOI-1615 (a magnitude 9.7 target) has a denser background of tens of magnitude 17-18 stars within 100 arc-seconds, with one 13.5  magnitude star at 50 arcseconds. The centroid motion of eclipses of all the targets is limited below one arcsecond, except for TOI-1153, which has a scatter of about $\SI{2}{arcsec}$. The contribution from these distant sources is therefore negligible. For information about possible close-in companions identified through speckle imaging, we invite the reader to consult the corresponding subsections of Sect.~\ref{sec:results}.

Before fitting, we phased the light curves from all available sectors to make the data set more manageable. We used the linear ephemeris found in the headers of the Threshold Crossing Event files of the TESS Data Products and number of bins to retain the original cadence. In the case of TOI-1788, the light curve was dominated by variability, with a frequency different from the orbital frequency as can be seen in the bottom panel of Fig. \ref{fig:1788lc}. We used {\sc Period04} developed by \citet{2005CoAst.146...53L} to remove this contribution. First, we masked the eclipses and then computed the Fourier spectra to obtain the dominant frequency. Next, we fit the sine of this frequency and computed spectra of the residuals. In this manner, we iteratively identified six main frequencies down to a relative flux amplitude of $0.0004$, obtaining a fit of the following form: 
\begin{equation}
F=F_0+\sum A_i 2 \pi (f_i t+\Phi_i)\,,
\end{equation}
where $A_i$ is the amplitude of the contribution of frequency $f_i$ with phase shift $\Phi_i$.

\subsection{Ond\v{r}ejov Observatory spectroscopy}

We used the Ond\v{r}ejov Echelle Spectrograph (OES) installed on the Perek 2-m telescope operated by the Astronomical Institute of the Czech Academy of Sciences. OES is a fiber-fed \'echelle spectrograph with a resolving power of $R\approx \num{50000}$ at $\SI{500}{nm}$, covering the wavelength range of about 380 to 900 nm \citep{2020PASP..132c5002K}. The wavelength calibration was performed with ThAr lamp frames taken during the observing night. We observed the stars over a total of 57 nights. The typical exposure time was between 1800 and 3600 seconds. The signal-to-noise ratio (S/N) achieved ranged from 15 to 70. 

The raw spectra were reduced with custom-prepared scripts using classical IRAF routines. The spectra were then corrected for instrumental shifts using telluric lines. The final product is a 1D merged spectrum from which the radial velocities were determined with IRAF's fxcor routine.

\subsection{Tautenburg Observatory spectroscopy}

The Tautenburg Coud\'e Echelle spectrograph (TCES) is installed on the 2m Alfred-Jensch telescope operated by the Th\"uringer Landessternwarte Tautenburg. 
The instrument achieves a resolving power of $R=\num{67000}$ with a slit-width of 1.2 arcsec. TCES has three cross-disperser grisms. For the observations, we used the so-called VIS grism, which covers the wavelength range from 460 to 730 nm.  TCES is located in a temperature-stabilized room in the basement of the dome. The light from the star is injected into the spectrograph via a Coud\'e mirror train from the telescope.
A temperature-stabilized $\mathrm{I}_2$ vapor absorption cell placed in front of the slit provides a dense grid of absorption lines in the wavelength range from 500 to 640 nm.
These iodine lines were used to determine the instrumental shift during the observations. 
Standard IRAF routines are used to subtract the bias, flat-field the spectra, remove the scattered light, extract the spectra, and wavelength-calibrate them.  Radial velocity measurement was performed using Velocity and Instrument Profile Estimator \citep[VIPER;][]{2021ascl.soft08006Z}.

\subsection{FIES spectroscopy}

We spectroscopically monitored TOI-416 with the FIbre-fed Echelle Spectrograph \citep[FIES;][]{1999anot.conf...71F,2014AN....335...41T} mounted at the 2.56m Nordic Optical Telescope (NOT) at Roque de los Muchachos Observatory (La Palma, Spain). We acquired seven spectra using the FIES \texttt{Med-Res} fiber (R\,=\,45\,000) at seven different epochs between January and February 2019, as part of the observing programs 58-024, 58-111, and 58-301 (PIs: Gandolfi, Hatzes, Deeg, respectively). We set the exposure time to 900-1500\,s based on sky conditions and observing schedule constraints. Following \citet{2010ApJ...720.1118B}, we monitored the intra-exposure drift of the instrument by acquiring long-exposed (T$_\mathrm{exp}$\,$\approx$\,60-80\,s) ThAr spectra immediately before and after each target observation. The FIES spectra were reduced using standard routines, as described in \citet{2013A&A...557A..74G,2015A&A...576A..11G}. Relative radial-velocity measurements were extracted using multi-order cross-correlations against the stellar spectrum with the highest S/N.

\subsection{SONG spectroscopy}
We also employed the 1m Hertzsprung SONG telescope 
\citep{2017ApJ...836..142G} and its high-resolution \'echelle spectrograph 
located  at Observatorio del Teide. The seven spectra obtained on seven 
consecutive nights starting January 28 have spectral resolutions of 
$\approx 80\,000$  and exposure times of \SI{45}{min}. We used the procedures outlined by \cite{2017ApJ...836..142G} to extract the spectra. Bad pixels were removed, and the spectra were normalized. Subsequently, 
we used a template spectrum ($T_{\mathrm eff} =6300$~K, $logg=4.5$ 
[Fe/H] =0) from the Phoenix library \citep{2013A&A...553A...6H} to 
create cross-correlation functions (CCFs) from which we derived the RVs 
used in this work.

\subsection{TRES spectroscopy}

We used the Tillinghast Reflector Echelle Spectrograph (TRES) \citep{Furesz} on the 1.5-m Tillinghast Reflector telescope at the Fred Lawrence Whipple Observatory (FLWO) in Arizona. TRES is a fiber-fed \'echelle spectrograph with a resolving power of $R\approx 44\,000$ operating in the  $\SIrange{390}{910}{nm}$ wavelength
range. Ten spectra were obtained from TOI-1143 during the period from November 2019 to February 2020. The exposure time ranged from 200 - 600
seconds, resulting in a S/N per resolution element of 16-43. The
spectra were processed using a custom pipeline as described in \citet{Buchhave2010} using ThAr spectra obtained before and after observations as a calibration source. A multi-order spectral analysis was then performed. In short, each observation was cross-correlated against the strongest S/N spectrum, order-by-order, to obtain radial velocities.

\subsection{WIYN imaging observations}
We observed TOI-1143, TOI-1153, and TOI-1788 using the NN-EXPLORE Exoplanet Stellar Speckle Image (NESSI) \citep{2018PASP..130e4502S} at the WIYN 3.5~m telescope on Kitt Peak, Arizona, USA.  NESSI is a dual-channel instrument that obtains simultaneous
speckle measurements in two filters using a pair of high-speed electron-multiplying CCDs.  For each of our target stars, we obtained 9000 40~ms
frames through filters with central wavelengths $\lambda_c = 562$ and
832~nm and band-pass widths of $\Delta\lambda = 44$ and 40~nm, respectively.
The readout of each camera was restricted to a $256\times256$~pixel region,
resulting in a $4.6\times4.6$~arcsecond field of view.  In addition to the
TOI targets, we obtained similar sets of 1000 speckle frames toward nearby
single stars in order to calibrate the observational point spread function (PSF).

We reduced the speckle data using a pipeline described in \citet{2011AJ....142...19H}.  Among the pipeline products is a reconstructed image of the field
around each TOI in each filter.  We identified any companion sources and
measured contrast curves using these images.  Although the reconstructed
image covers a larger area, we have limited our contrast curves to an outer
radius of 1.2~arcseconds to avoid the decorrelation of speckle patterns at
wider separations.  Another pipeline product is a model fit to the power
spectrum, which is derived from the Fourier transform of the mean of the
autocorrelation function of each speckle frame.  When a companion star is
identified, a model fit to this power spectrum provides a measurement of
the relative astrometry and photometry of the two sources. There was only one system with an identified companion, TOI-1143, which is shown in Fig. \ref{fig:1143im}.

\section{Spectral-energy-distribution fitting}
To obtain the physical characteristics of the systems, we need some information about the primary stars. The measured radial-velocity amplitudes suggested the companions to be low-mass stars. The stellar parameters of the primaries were obtained by fitting their SEDs with {\sc Ariadne} \citep{Ariadne} -- at first under the assumption that the flux contribution of the secondary is negligible. {\sc Ariadne} downloads available broad-band photometry,\footnote{ALL-WISE \citep{WISE}, APASS DR9 \citep{APASS}, Pan-STARRS1 \citep{PanSTARS}, SDSS DR12 \citep{SDSS}, 2MASS \citep{2MASS}, Tycho-2 \citep{Tycho}, ASCC \citep{ASCC}, GLIMPSE \citep{GLIMPSE}, GALEX \citep{GALEX}, Strömgen Photometric Catalog \citep{ernst}, Gaia DR2\citep{Gaia1, Gaia2}.} a subset of which is presented in Table~\ref{tab:SED-data}, and fits it with the atmospheric models of
Phoenix v2 \citep{Phoenix}, \citet{Kurucz}, \citet{CK}, BT-Settl \citep{BTSettl}, BT-NextGen \citep{NextGen,BTSettl}, and BT-Cond \citep{BTSettl}. From these, the first four mentioned were used because BT-Settl, BT-NextGen, and BT-Cond are identical above $T>\SI{4000}{K}$. To create model grids, atmospheric models corrected for distance, and interstellar absorption are convolved with the respective photometric spectral response functions to obtain model observed fluxes. 

We fit for effective temperature ($T_{\rm{eff}}$), radius ($R$), distance ($D$), surface gravity ($\log g$), metallicity ([Fe/H]), and line-of-sight extinction ($A_V$). For $T_{\rm{eff}}$ and $\log g$ priors, we used the TIC v8 catalog \citep{TICv8} values. For [Fe/H], $R$, $D,$ and $A_V,$ we used the default {\sc Ariadne} priors.  For [Fe/H], {\sc Ariadne} uses a Gaussian prior of $\mathcal{N}(-0.125,0.234)$ based on data from the spectroscopic RAVE survey \citep{RAVE},\footnote{ Estimated metallicities from {\sc Ariadne} are comparable with those derived by spectral-reduction pipelines in ExoFOP. Moreover, metallicity only enters our model through pass-band luminosities via the atmospheric tables where the influence is known to be negligible.} $D$ is drawn from the Bailer-Jones distance estimate from Gaia \citep{BJ}, the prior for $R$ is in the interval from 0.5 to 20 \rs, while the $A_V$ prior is limited to the maximum line-of-sight extinction from the updated SFD Galactic dust map \citep{SDF}. Parameters obtained using {\sc Ariadne} under the one-star assumption are in Table~\ref{tab:ariadne}. We stress that these parameters were obtained assuming that the stars are single. Nevertheless, we see good agreement between measured and modeled fluxes, so we used these values as good estimates of primary star properties within the binary systems. Later, we subtracted model fluxes of the secondaries to refine the primary star parameters. These fits can be seen in Fig. \ref{fig:SED} in the appendix.

\begin{table}
\begin{center}
\caption{Astrometric and photometric data used for SED fitting.}
\label{tab:SED-data}
\resizebox{\columnwidth}{!}{
\setlength{\tabcolsep}{2pt}
\begin{tabular}{c|ccccc}
TOI     &       416             &       1143            &       1153            &       1615            &       1788\\  
\hline
$\alpha_{2000}$ [h m s] & 03 15 25.1 & 12 10 09.3   & 12 10 46.0    &23 24 41.9 &10 40 55.0\\
$\delta_{2000}$ [\dg{} ' "] & -02 10 31 &+77 21 08& +85 42 18 &+77 37 37&+35 54 48\\
$\mu_\alpha$ [mas/yr]    & 29.76(9) & -36.30(11)    & -14.19(6)     &12.59(5)   &-48.38(8)\\
$\mu_\delta$ [mas/yr]    & -6.86(8) &4.19(8)        & -2.27(5)      &2.74(5)    &-72.31(7)\\
$D$ [pc]        &  132.0(8)     &  192.0(20)        &147.1(6)       &260.5(21)  &191.5(24)    \\
\hline
BT&     9.423(21)       &       9.706(22)       &       9.250(18)       &       10.293(27)      &       11.602(68)\\
B&      9.245(8)        &       9.599(36)       &       9.217(14)       &       10.388(92)      &       11.468(14)\\
BP&     8.921(3)        &       9.262(3)        &       8.900(3)        &       10.064(3)       &       10.970(4)\\
VT&     8.828(16)       &       9.138(17)       &       8.821(14)       &       10.008(27)      &       10.726(53)\\
y&      8.770(20)       &       --              &       8.777(14)       &       10.09(11)       &       10.757(32)\\
V&      8.929(4)        &       9.077(2)        &       8.773(2)        &       9.975(4)        &       10.602(17)\\
G&      8.6604(28)      &       9.006(3)        &       8.701(3)        &       9.912(3)        &       10.637(3)\\
RP&     8.2298(38)      &       8.581(4)        &       8.347(4)        &       9.639(4)        &       10.135(4)\\
T&      8.282(6)        &       8.633(6)        &       8.405(6)        &       9.6871(6)       &       10.188(6)\\
J&      7.776(24)       &       8.119(24)       &       7.965(29)       &       9.326(23)       &       9.628(24)\\
H&      7.557(46)       &       7.874(20)       &       7.798(31)       &       9.229(27)       &       9.347(35)\\
K&      7.473(21)       &       7.829(16)       &       7.738(16)       &       9.159(21)       &       9.239(23)\\
W1&     7.383(36)       &       7.778(26)       &       7.706(30)       &       9.126(24)       &       9.160(23)\\
W2&     7.458(20)       &       7.818(20)       &       7.733(23)       &       9.132(20)       &       9.188(20)\\

\end{tabular}}
\end{center}
\end{table}

\begin{table}
\begin{center}
\caption{Results of initial {\sc Ariadne} SED fitting ignoring the secondary components.}
\label{tab:ariadne}
\resizebox{\columnwidth}{!}{
\setlength{\tabcolsep}{2pt}
\begin{tabular}{c|ccccc}
TOI &   416&    1143&   1153&   1615&   1788\\
\hline
$T_\mathrm{eff}$ [K]       &6238(70)    &6242(65)       &6691(80)       &7120(80)       &5742(34)\\
$\log g$                       &4.02(40)        &4.37(30)       &3.47(43)       &4.11(30)       &4.03(60)\\
$[\mathrm{Fe}/\mathrm{H}]$ &-0.18(10)   &-0.23(12)      &-0.06(7)       &-0.20(15)      &0.02(6)\\
$D$ [pc]                       &130.92(17)      &187.30(12)     &145.61(12)     &254.8(12)      &192.19(45)\\
$R$ [\rs]                      &1.840(25)       &2.234(25)      &1.734(22)      &1.533(22)      &1.264(17)\\
$A_V$                      &0.09(4)         &0.07(4)    &0.09(6)        &0.04(5)        &0.03(1)\\
\end{tabular}}
\end{center}
\end{table}

\section{Light curve and radial-velocity fitting}
\label{sec:Phoebe}

To model the light curves, we used the {\sc Phoebe} software developed by \cite{2016ApJS..227...29P}, which is an eclipsing binary modeling code with Python interface\footnote{Freely available to download along with documentation and many examples at \url{http://phoebe-project.org/}.}  using the {\sc Phoebe}~2.4 version. The astrophysical background of eclipsing binary stars and their modeling is described in detail in the book by \citet{2018maeb.book.....P}. We generally followed the guide by \citet{2020ApJS..250...34C} for the current version. We are, however, using {\sc Phoebe} mainly as a wrapper instead, using the {\sc ellc} forward model by \citet{2016A&A...591A.111M} describing the stars as ellipsoids and using Gauss-Legendre integration over projected ellipses to obtain measured fluxes. Pass-band fluxes and logarithmic limb-darkening coefficients were sourced from Phoenix atmosphere tables by \citet{2013A&A...553A...6H} to cover even the coolest companions. 

In all cases, we used synchronously rotating stars in our models to describe the ellipsoidal shape of the stars. In case of small deviations from sphericity, the introduced error in stellar parameters is minimal. Bigger deviations are driven by more massive secondaries on close-in orbits, where synchronous rotation is expected. To illustrate the level of the possible errors introduced by this assumption, we estimated relative differences between the smallest and the biggest radius of curvature of tidally deformed stars. This would be the radii driving the shape of the light curve during eclipse for grazing eclipses with a tidal bulge aligned with the secondary and with a $\SI{90}{\degree}$ lag. Using values from \cite{2012Ap&SS.338..127P} for a mass ratio of $q=0.4$ and fill-out factor of $F=0.4$ (corresponds to $R_1/a\doteq 0.16$) as a worst case scenario for our systems, we see a difference of only $1.8\%$. This upper estimate of the introduced error is of the same magnitude as the uncertainty in the primary radii themselves. In case of systems with precisely determined parameters - TOI-416 and TOI-1143 - this estimate is closer to the case with $q=0.1$, $F=0.2,$ leading to a difference of only $0.23\%$. To obtain the proper posterior distributions of model parameters, we used the {\sc EMCEE} Markov chain Monte Carlo sampler by \citet{2013PASP..125..306F}.

\subsection{Preliminary step}
In some of our systems, the effect of Doppler beaming and boosting is clearly visible (see \citealt{2007ApJ...670.1326Z} for the description of Doppler beaming in a binary star context).  The current version of {\sc Phoebe} does not support this effect; therefore, we proceeded to fit the light curves outside the eclipses by the following function:
\begin{equation}
\label{eq:beam-fit}
F(\Phi)=F_0-A\cos(2\pi\Phi)+B\sin(2\pi\Phi)-C\cos(4\pi \phi) +D\sin(a\pi\Phi)\,,
\end{equation}
which describes the total flux $F$ as a function of an orbital phase $\Phi$. The sum consists of a constant term $F_0$, a reflection term with amplitude $A$, a relativistic beaming and boosting term of amplitude $B$ and geometric ellipsoidal variation of amplitude $C$. The coefficient $D$ should be zero and was added as a further mean to estimate errors of the other parameters. This description was shown by \citet{2011MNRAS.415.3921F} as a good means of describing out-of-eclipse variability in close binaries. We note that the boosting term is proportional to the radial velocity, so the above description is valid for circular orbits only. The constant of proportionality can, in principle, be derived from the spectra of the star and the detector pass band. For now, we use it as a free parameter instead.

\begin{table}[]
 \centering 
 \label{tab:beam-fit}
\caption{Fits of light curves out of eclipse according to Equation~\ref{eq:beam-fit}.}
\begin{tabular}{c|cccc}
System&$A\cdot 10^5$&$B\cdot 10^5$&$C\cdot 10^5$&$D\cdot 10^5$\\
\hline
TOI-416     &-1.5$\pm$6.1   &10.9$\pm$1.4   &11.0$\pm$2.0 &2.3$\pm$3.7\\
TOI-1153    &10.7$\pm$4.3   &24.3$\pm$2.5  &30.7$\pm$2.6&1.1$\pm$2.8\\
TOI-1615    &16.7$\pm$5.3   &25.6$\pm$9.2  &36.4$\pm$9.7&3.5$\pm$5.8\\
TOI-1788 $^a$    &-10.1$\pm$1.6  &0.1$\pm$0.2   &-7.5$\pm$1.5& $\pm$ \\
\end{tabular}

\footnotesize $^a$ Only one sector available; errors are from the least-squares fit and are thus grossly underestimated.
\end{table}

We performed separate fits for every sector and then computed averaged values of the coefficients presented in Table~\ref{tab:beam-fit}. We omit the $F_0$ term from the table. It is only a shift of a few $10^{-5}$  in normalization,  given by the difference between the flux used to normalize the light curve initially and the mean flux of the stars in the ellipsoidal model. The mean normalized flux of the system is a free parameter in our subsequent model. Signs in Equation~\ref{eq:beam-fit} were chosen so that the respective coefficients, $A$, $B$, and $C,$ would be positive. Parameter errors of the least-squares fits were around $(0.5-1.5)\cdot 10^{-5}$, while the scatter of the values between sectors was larger. We attribute this discrepancy to the instrumental systematics in the light curves and to the detrending algorithm used by the {\sc PDC-SAP} pipeline. The values of $D$ are consistent with zero. We see that in the case of TOI-1788, the value of $B$ is quite negative - this is a consequence of the smoothing performed for this system to remove rotational variability, and thus we cannot expect the residual out-of-eclipse variability to be of an astrophysical nature. As a result, for TOI-1788 we only model the primary eclipse. For the other systems, we subtracted the contribution of beaming and boosting from the data such that {\sc PHOEBE} can model the other two effects directly. We excluded TOI-1143 from this quick analysis because of its eccentric orbit.

\subsection{Main step in PHOEBE}
As the next step, we jointly modeled the radial velocities from all available instruments using the period derived from the eclipses.  We used the following set of variables as the parameters to be estimated: the projected semi-major axis of the primary around the barycenter ($a_1 \sin i$), the eccentricity projections ($e\sin \omega$, $e\cos \omega$), and the systemic velocities ($\gamma_i$) of each data set and their respective factors of error underestimation ($\sigma_i$). From this, the primary's radial-velocity semi-amplitude ($K_1$) and the mass function ($f(M)$) can be computed using the period of the system ($P$) and Newton's gravitational constant ($G$) as follows:
\begin{equation}
K_1=\frac{2\pi}{P} \frac{a_1 \sin i}{\sqrt{1-e^2}}\,,\quad f(M)= \frac{4 \pi^2}{P^2}\frac{(a_1 \sin i)^3}{G}\,.
\end{equation}

Next, we fit the light curves solving for the ratio of stellar radii ($R_2/R_1$), inclination ($i$), ratio ($(R_1+R_2)/a$) of the sum of stellar radii over the semi-major axis, primary pass-band luminosity ($L_\mathrm{pb,1}$), and ratio of secondary and primary effective temperature ($T_2/T_1$) as well as the projection of $e\cos i$ if a secondary eclipse was present. Here, we fixed the primary radius ($R_1$) and temperature ($T_1$) according to the {\sc Ariadne} fits, and $a_1 \sin i$ and eccentricity projections according to the radial-velocity fit.  We used initial values suggested for main-sequence stars by \citet{2013ApJS..208....9P} using the TESS Input Catalogue parameters as a guide for the primary, and  we manually adjusted the inclination to match the depth of the eclipses.  From these values, the semi-major axis of the binary and mass ratio can be computed as
\begin{equation}
a=R_1 \frac{1+\left(\frac{R_2}{R_1}\right)}{\left(\frac{R_1+R_2}{a}\right)}\,,\quad q=\frac{\left(a_1 \sin i\right)}{a \cdot \sin i - \left( a_1 \sin i \right)}\,,
\end{equation}
which yields the masses of the components using the third Kepler's law.

As a final step, we refined the parameters by running the radial velocity and light-curve fit together. We also\textbf{\textbf{}} attempted to fit for the albedo of the secondary ($A_2$) and the gravity brightening coefficient of the primary ($\beta_1$), but we did not achieve conclusive results. We note that the simultaneous solution of radial velocities and light curves often converged to incorrect values of eccentricity projection, $e\sin \omega$ and $a_1 \sin i$, with only small visual differences in light curves but unsatisfactory fits in radial velocities. We believe this is the consequence of light-curve domination of the solution given by the order-of-magnitude difference in the number of photometric and RV points used in the fit.

\subsection{Correction for binarity}
After successfully fitting the combined radial-velocity and photometry data with {\sc Phoebe}, we computed magnitude corrections from fractional light in the photometric channels to obtain magnitudes of the primary star alone. We subsequently used these to perform new {\sc Ariadne} SED fits. If the results were similar to the previous values, we performed the following simple corrections for the parameters. Fixing the previously fit $T_2/T_1$, $R_2/R_1$ and $(R_1+R_2)/a$, we obtain corrected values for secondary star temperature and radius and orbital semi-major axis as a simple scaling:
\begin{equation}
    T_2=\frac{T_1}{T_1^{\rm old}} T_2^{\rm old}\,,\quad
    R_2=\frac{R_1}{R_1^{\rm old}} R_2^{\rm old}\,,\quad
    a= \frac{R_1}{R_1^{\rm old}} a^{\rm old}\,.
\end{equation}
Radial-velocity-dependent parameters are fixed, so for the mass ratio we have
\begin{equation}
\frac{aq}{1+q}= a_1 =\mathrm{const.} \quad \xrightarrow{} \quad
    q=\frac{a^{\rm old}}{a} \frac{q^{\rm old}}{1+q\frac{a-a^{\rm old}}{a}}\approx \frac{R_1^{\rm old}}{R_1} q^{\rm old}\,.   
\end{equation}
For the masses, we thus have the following correction from Kepler's third law:
\begin{equation}
M_1=\left(\frac{R_1}{R_1^{\rm old}}\right)^3 M_1^{\rm old}\,,\quad M_2=\left(\frac{R_1}{R_1^{\rm old}}\right)^2 M_2^{\rm old}\,.
\end{equation}
These equations also serve to estimate the uncertainty of these parameters due to the uncertainty of the primary radius, which was fixed during the {\sc Phoebe} run.

For cases where the binarity corrections for $R_1$, $T_1$ were larger than the uncertainties on the values themselves, we made a new {\sc Phoebe} photometry model to accommodate this. The main reason is that the temperature change can influence coefficients of limb darkening and pass-band luminosities, which can change the shape and depth of the eclipse in the forward model necessitating a change in inclination $i$ and ratios $R_2/R_1$ and $(R_1+R_2)/a$. 

\section{Results}
\label{sec:results}
\subsection{TOI-416}
\begin{figure}[h!]
    \centering
    \includegraphics[width=\linewidth]{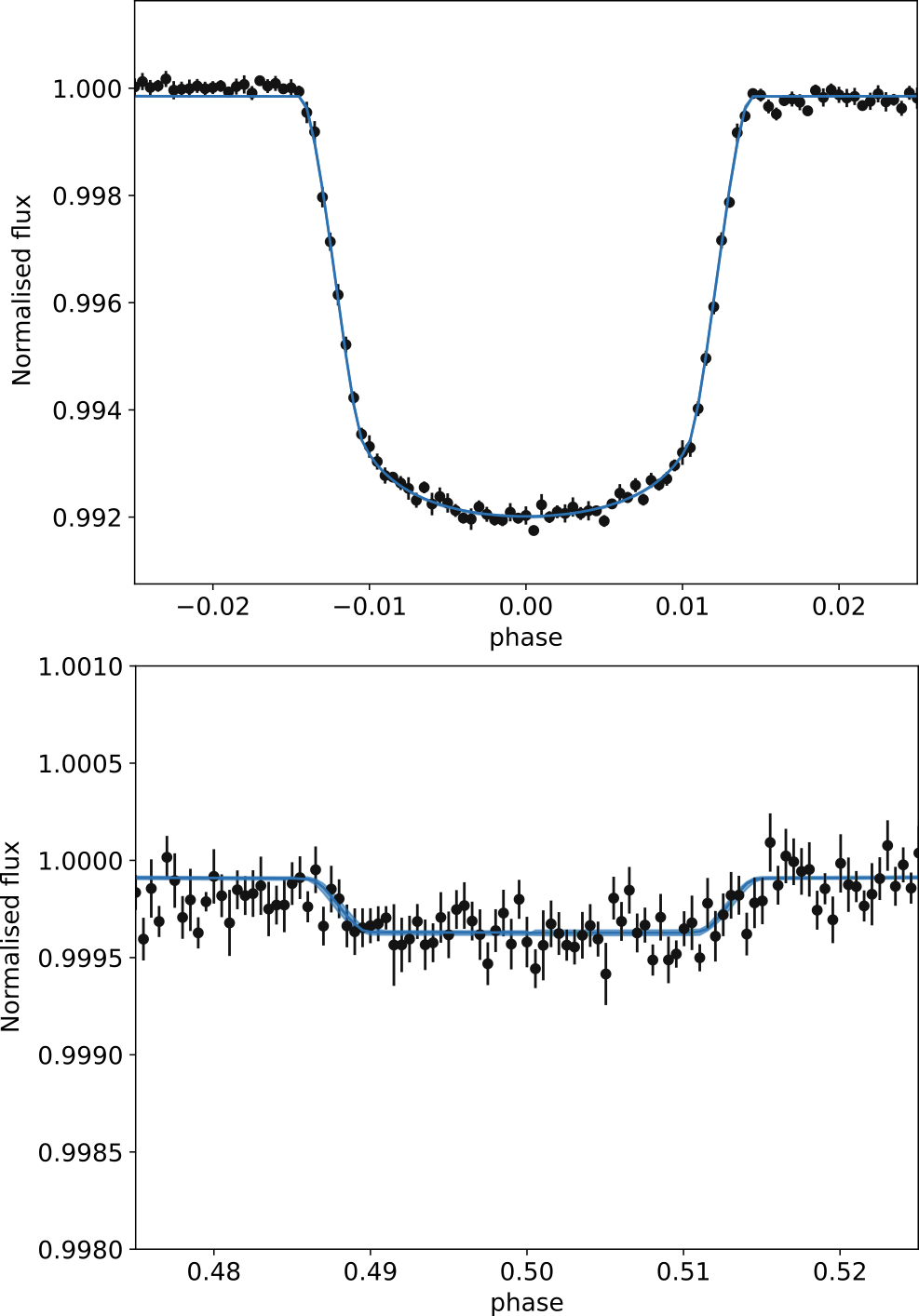}
    \caption{Light curve of TOI-416. We can see the total primary eclipse in the upper panel and shallow secondary eclipse in the bottom panel.}
    \label{fig:416lc}
\end{figure}

\begin{figure}[h!]
    \centering
    \includegraphics[width=\linewidth]{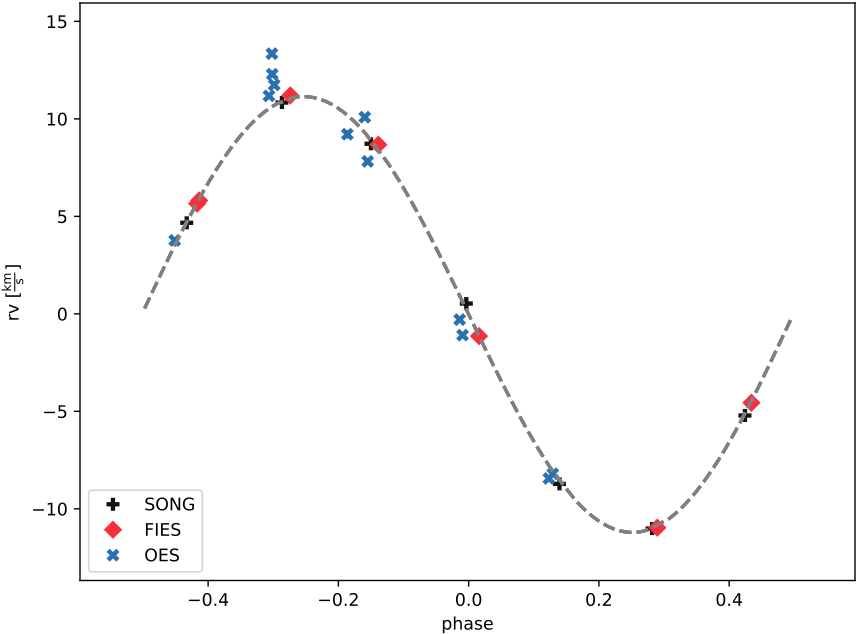}
    \caption{Radial velocity curve of TOI-416 obtained by OES, SONG, and FIES corrected for the instrumental zero points with the best-fit solution shows a circular orbit of the system. Errors of the data points are smaller than the markers.}
    \label{fig:416rv}
\end{figure}

The seventh-magnitude star HD 20267 (BD-02 583, TIC 49899799) in the northern part of Eridanus was identified as an exoplanetary candidate, TOI-416, in January 2019. It was also recovered in an eclipsing binary study by \cite{2022ApJS..258...16P} searching for eclipsing binaries in TESS. The star shows total primary eclipses with secondary eclipses also visible on a nearly circular seven-day orbit, as can be seen in Fig.~\ref{fig:416lc}. There is no significant contaminant in the surrounding field. The Gaia astrometric solution yields a renormalized unit weight error (RUWE) of 1.12, and there are no detected companions within  $0.2 "$ (which is about $\SI{30}{au}$ at the distance of the main source) according to speckle imaging at SOAR by A. Tokovinin available via ExoFOP archive. In our analysis, we used our 12 radial-velocity measurements from the Ond\v{r}ejov Observatory, seven measurements from the SONG node on the island of Tenerife by S. Albrecht via the ExoFOP archive, and seven measurements by FIES. In Fig.~\ref{fig:416rv}, we can see a circular orbit with an amplitude of about $\SI{11}{km/s,}$ signalling a very low-mass stellar companion. 

Our SED fit, corrected for the light of the secondary, yielded a temperature of $T_1=\SI{6330(60)}{K}$, radius of $R_1=\SI{1.82(2)}{R_\odot}$, surface gravity of $\log g=\SI{4.02(5)}{}$, metallicity of $[\rm {Fe/H}]=-0.18(10)$, and an isochrone mass of $M_1=\SI{1.31(7)}{M_\odot}$. We note that these values are comparable to Gaia spectroscopy and spectroscopy by A. Bieryla by TRES at FLWO, except for the temperature. Our temperature estimate is about $\SI{300}{K}$ higher than both TRES and Gaia. We also note that the projected rotational speed of the primary from the TRES spectral analysis of $v \sin i= \SI{10.3(5)}{km/s}$ is lower than the expected value for synchronous rotation $v_{syn}=\SI{13.1(2)}{km/s}$.

The resulting parameters from our fit can be seen in Table~\ref{tab:par}. Posteriors of the fit parameters were Gaussian, containing small covariances with the exception of $i$ and $(R_1+R_2)/a,$ which are tightly bound in anticorrelation by the duration of the eclipse. We also tried to fit for the bolometric (Bond) albedo of the secondary ($A_2$) and bolometric gravity darkening coefficient of the primary ($\beta_1$), without success. We thus kept them at suggested values: $\beta_1=\beta_2=0.32$ and $A_1=A_2=0.6$. We can see that the orbit of the star is indistinguishable from a circular one. The binary consists of a $\SI{1.36(7)}{M_\odot}$ primary (in good agreement with the {\sc Ariadne} fit) evolving from the main sequence and a secondary with a mass of just $\SI{0.131(8)}{M_\odot}$ on the main sequence of spectral class M5V. Our {\sc Ariadne}-MIST estimate for the age of the primary and thus the age of the system of $\SI{2.9(5)}{Gyr}$ provides enough time for the circularization of the orbit. The slow rotation speed of the primary can be explained by the evolutionary expansion of the envelope of the star, which slows down the rotation from the previous synchronized state.

\subsection{TOI-1143}
\begin{figure}[h!]
    \centering
    \includegraphics[width=\linewidth]{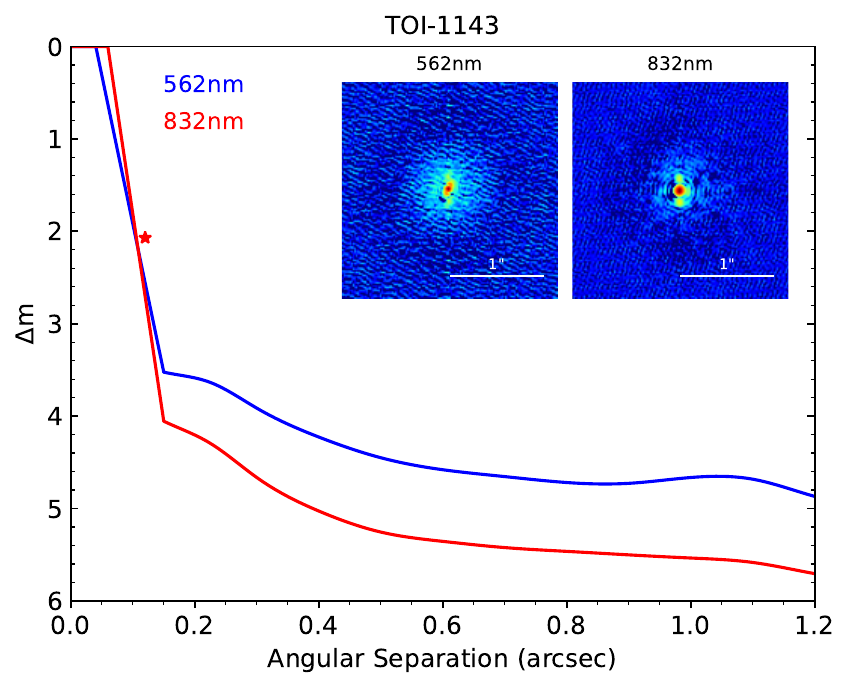}
    \caption{Contrast curves for TOI-1143 showing the detection limit of NESSI speckle imaging in the 562~nm (blue curve) and 832~nm filters (red curve).  The star symbol indicates the separation and contrast of the detected companion.  A reconstructed image for each filter is shown in the upper right, with TOI-1143 in the center of the field.  The companion appears as a pair of close-by sources lying roughly north and south of TOI-1143.  The double image is due to an ambiguity in the image reconstruction.  These images are oriented with north at the top and east to the left.}
    \label{fig:1143im}
\end{figure}

The northern star of ninth magnitude, BD+78 407 (TIC 160268701), was flagged as a possible exoplanetary candidate by TESS in August 2019. The light curve shows an 11.2-day eccentric orbit with a shallow secondary minimum around phase 0.4. A search of the Gaia catalog reveals no close stars significantly contaminating the light of the target. However, the star shows quite a big astrometric excess error with $\mathrm{RUWE}=5.05,$ hinting at a possible close companion. This star was resolved by NESSI at a separation of $0.12\arcsec$ and a magnitude difference of ${\Delta}m=2.07(10)$ in both V and I filters, as can be seen in Fig. \ref{fig:1143im}. The position angle of the companion star in this instance is
ambiguous by 180 degrees, meaning its true position angle could be either $\SI{4}{\degree}$ or $\SI{184}{\degree}$. We adopted a third-light ratio of $12.9(10)\%$ in the photometric fit, and we corrected the radius of the star from {\sc Ariadne} by the corresponding amount. The nature of this star is discussed later. We had 40 radial-velocity measurements from Ond\v{r}ejov and ten from TRES at our disposal. To obtain a more precise linear ephemeris, we also used long cadence {\sc TESS} observations from Sector 14 on top of short cadence sectors 47, 48, 53, and 60.

After the correction for the binary companion and the contaminant we obtained a temperature $T_1=\SI{6300(50)}{K}$, radius $R_1=\SI{2.069(18)}{R_\odot}$, $\log g=\SI{3.85(5)}{}$ and metallicity $[\mathrm{Fe/H}]=-0.21(10)$. This is slightly hotter than the spectroscopic fit by L. A. Buchhave using NOT-FIES data with $T=\SI{6100(50)}{K}$, while the surface gravity and metallicity are in good agreement. They also reported a value of projected rotational velocity of $v\sin i=\SI{11.1(5)}{km/s,}$ implying a rotational period of up to $P_{rot}\doteq \SI{9.4}{day,}$ which is shorter than the orbital period. However, due to the eccentricity of the orbit, the expected pseudo-synchronized rotational period of the primary \citep{1981A&A....99..126H} is about $P_{rot}^{ps}=\SI{7.5}{day}$, which is greater than the observed one.

\begin{figure}[h!]
    \centering
    \includegraphics[width=\linewidth]{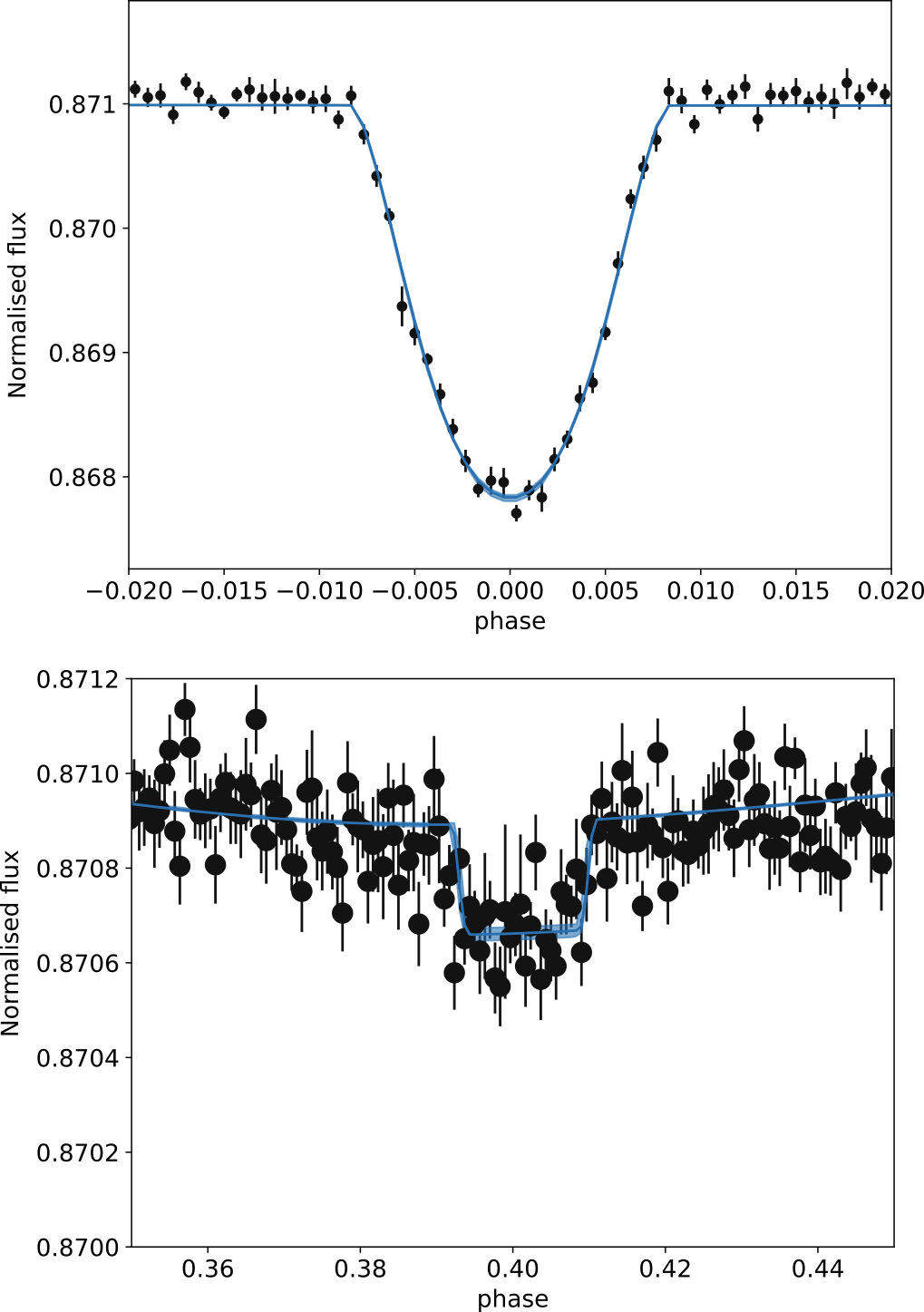}
    \caption{Light curve of TOI-1143. We can see the total primary eclipse in the top panel and shallow secondary eclipse offset from the phase 0.5 by its eccentric orbit in the bottom panel.}
    \label{fig:1143lc}
\end{figure}

\begin{figure}[h!]
    \centering
    \includegraphics[width=\linewidth]{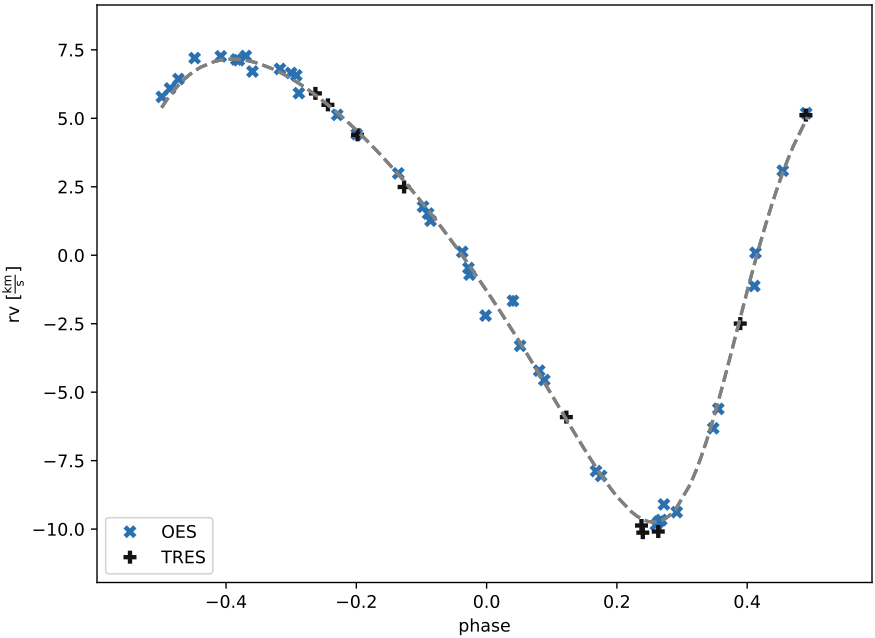}
    \caption{Radial velocity curve of TOI-1143 obtained by OES and TRES corrected for instrumental zero point with our best-fit solution on an eccentric orbit. Errors of the data points are smaller than the markers.}
    \label{fig:1143rv}
\end{figure}

Our photometric fit in Fig.~\ref{fig:1143lc} models the observed light curve nicely using the suggested gravity-brightening-coefficient values and albedos of the stars. The posteriors have only small covariances, with the exception of $i$ and $(R_1+R_2)/a$. The radial-velocity curve in Fig.~\ref{fig:1143rv} reveals the eccentricity of the orbit and low mass of the stellar companion. Furthermore, two data points near the primary eclipse are outlying suggesting presence of the Rossiter-MacLauglin effect. However, the expected amplitude is well under $\SI{100}{m/s,}$ and latter point happens after the eclipse is finished. We therefore discarded these points in our analysis. The parameters of the system can be found in Table~\ref{tab:par}. The primary star is an evolved sub-giant star. \citet{2013ApJS..208....9P} suggested a temperature about $\SI{700}{K}$ hotter and radius $23\%$ smaller for a main-sequence star of the same mass. The secondary has a mass of $M_2=\SI{0.118(5)}{M_\odot, which}$ is consistent with an M5.5 dwarf, although our temperature estimate of $T_2=\SI{3210(60)}{K}$ is about $\SI{300}{K}$ too high for a main-sequence star. This can be caused by an underestimation of the secondary albedo. However, the estimated temperature is for the irradiated face of the star just after periastron passage. It can be higher than expected in case the rotation of the secondary is synchronous and heat transfer not very effective.

Finally, we discus the nature of the companion detected in the imaging. The magnitude difference of 2.07 mag with respect to the primary in both V and I filters would suggest a star of the same temperature behind the binary system. In spite of that, the field surrounding TOI-1143 is quite sparse, containing only about 50 sources brighter than the imaged companion inside a one degree radius. If we consider a disk with a radius of $0.12"$, we would expect just $1:10^{-7}$ stars inside it at this stellar density. Therefore, it is highly unlikely that the imaged companion is just a star in the foreground or background. If we consider the errors of the contrast measurements $\Delta \mathrm{I}=0.1$ and $\Delta \mathrm{V}=0.15$, we obtain an error in the colour of $\Delta (\mathrm{V}-\mathrm{I})=0.18$. According to \citet{2013ApJS..208....9P}, for stars at the same distance (if we correct for the difference in radius of our primary and a main-sequence star of the same temperature), the contrast of $2.07$ in the I band would suggest a main-sequence star with a temperature around $\SI{5550}{K}$. Here, the companion would be redder by $0.18$ in V-I, which is not strongly ruled out by the imaging colour. We thus conclude that the imaged star is a bound companion in the system. 

The angular separation of the companion translates to a tangential separation of $\SI{23}{au}$ and minimal possible orbital period of $\approx \SI{25}{y,}$ assuming the currently observed separation is equal to twice the semi-major axis as an extreme scenario. This changes to $\approx \SI{100}{y}$ for $a=\SI{23}{au}$ equal to the tangential separation and $\approx \SI{200}{y}$ for $a=\SI{33}{au}$ for a circular orbit if the companion is seen at elongation and halfway between elongation and conjunction, respectively. This would imply a variability in systemic velocity of the central binary up to $\approx\SIrange{5}{10}{km/s}$ in the edge-on case and respective changes in eclipse timing of amplitude of tens of minutes up to an hour on a decadal to centennial timescale. We checked the residuals of the Ond\v rejov radial-velocity measurements spanning $\approx\SI{300}{d}$ by fitting a quadratic polynomial: $f(x)=a+bx+cx^2$. However, the constant comparison fit resulted in $\chi^2_{NDF}$ smaller by about 5\% preferring the non-detection of the companion. We placed $3\sigma$ upper limits on the coefficients $b<\SI{0.0021}{km/s/d}$ and $c<\SI{0.000033}{km/s/d^2}$. Assuming a circular orbit of period $P_3$, one can derive the relation for the observed reflex motion of the central binary:
\begin{equation}
    K_3=b^2\left(\frac{P_3}{2\pi}\right)^2+4c^2\left(\frac{P_3}{2\pi}\right)^4\,.
\end{equation}
Using our limits on $b$,$c$ and the minimal conceivable period $P_3=\SI{25}{y}$, we obtain the $3\sigma$ limit $K_3<\SI{140}{km/s}$. Our radial-velocity observations are thus unable to place any reasonable limit on the orbit of the third body in the TOI-1143 system.

For completeness, we also considered the option that the low-mass companion is orbiting the imaged companion. Assuming the third light $l_3=87\%$ and primary star radius $R_1=\SI{0.93}{R_\odot,}$ we obtained a radius for the secondary $R_2=\SI{0.16(1)}{R_\odot}$ and the sum of the masses of the components, $M_1+M_2=\SI{0.53(5)}{M_\odot}$. To explain the ellipsoidal variation, it is further necessary to use high-mass ratios around 1.7, leaving us with a primary star with a mass of just $M_2=\SI{0.20}{M_\odot,}$ which is clearly outside of range for both stars and white dwarfs. Therefore, we rule out this possibility.


\subsection{TOI-1153}
\begin{figure}[h!]
    \centering
    \includegraphics[width=\linewidth]{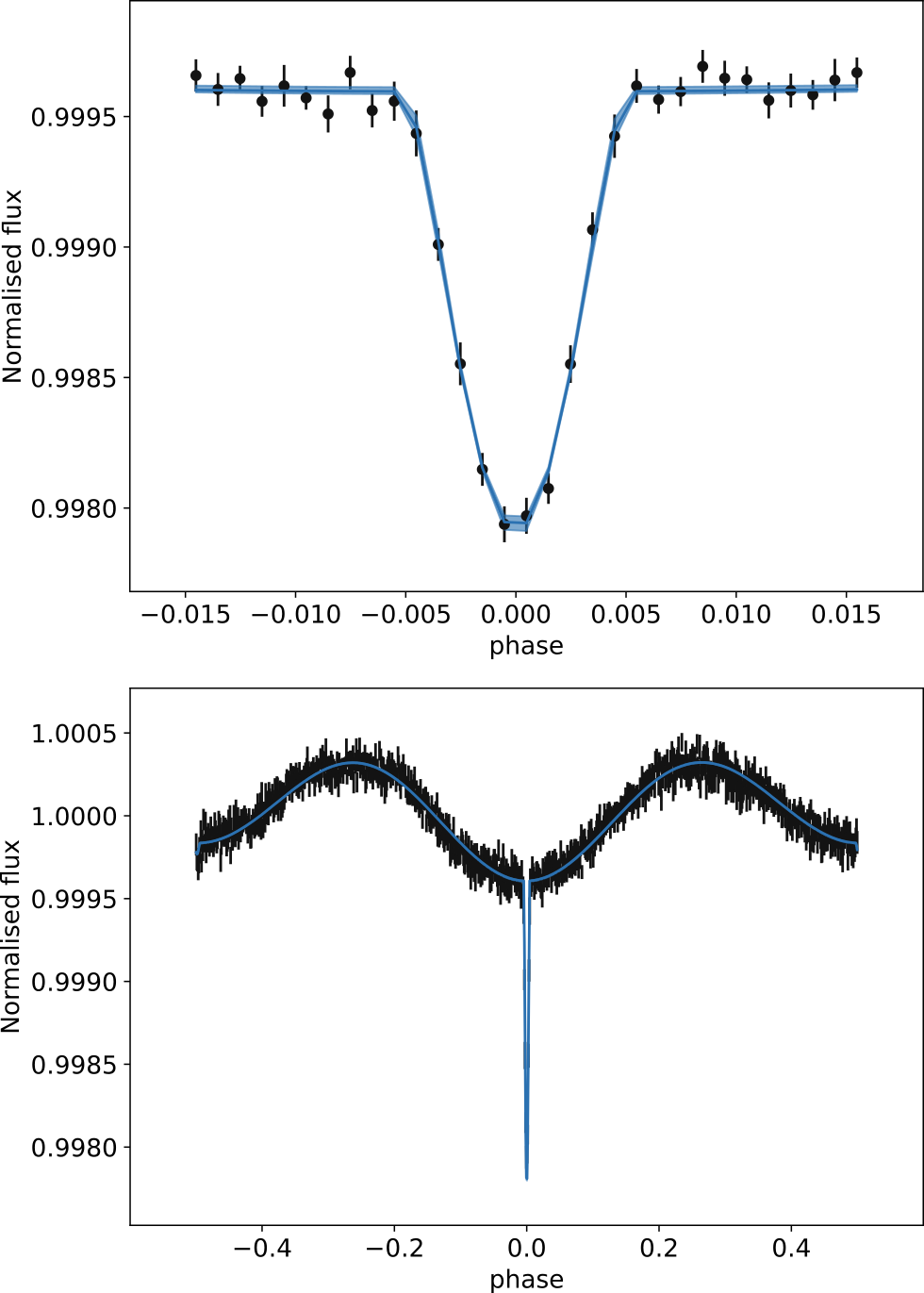}
    \caption{Light curve of TOI-1153. The primary eclipse is shown in the top panel, and the full phase curve showing significant ellipsoidal variation in the bottom panel. We do not detect conclusive evidence of secondary eclipses.}
    \label{fig:1153lc}
\end{figure}

\begin{figure}[h!]
    \centering
    \includegraphics[width=\linewidth]{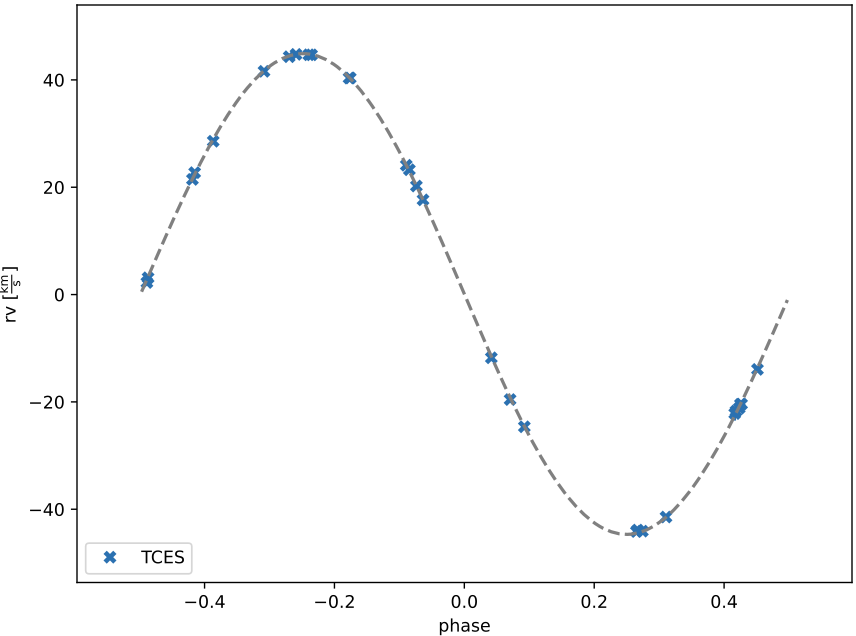}
    \caption{Radial velocity curve of TOI-1153 obtained by TCES after zero-point correction with our best-fit model showing a circular orbit. Errors of the data points are smaller than the markers.}
    \label{fig:1153rv}
\end{figure}

The ninth-magnitude star HD 106018 (BD+86 177, TIC154840461) in the northern part of Camelopardalis was recognized as a possible transiting exoplanetary candidate by TESS in September 2019. It was also recovered from the same data in an eclipsing binary study by \cite{2022ApJS..258...16P}. The light curve shown in Fig.~\ref{fig:1153lc} with a 6.0-day period features significant out-of-eclipse variations and a V-shaped primary eclipse, suggesting a binary star with grazing eclipses and ellipsoidal variability. The detection of an extremely shallow secondary eclipse in the binned data is tentative. There are no close companions detected in imaging with NESSI@WIYN by Mark Everett (with a sensitivity level of $3^m$ down to $0.12''$). It has  a low TIC-contamination ratio of only 0.00035 and a Gaia astrometric error of $\mathrm{RUWE}=1.057,$ further suggesting an uncontaminated binary system. We used 29 radial-velocity measurements by the Tautenburg Observatory, showing a circular orbit with a semi-amplitude of about $\SI{45}{km/s,}$ shown together with our fit in Fig.~\ref{fig:1153rv}.

The primary star has a temperature of $T_1=\SI{7000(130)}{K}$ and a radius of $R_1=\SI{1.669(22)}{R_\odot}$ after correction for the secondary. For comparison, the spectral fitting done by 
L. A. Buchhave using NOT-FIES yielded a temperature of $T_1=\SI{6700(50)}{K}$, $\log g=\SI{4.14(10)}{}$, metallicity of $[\mathrm{Fe/H}]=\SI{-0.04(8)}{}$ and high rotational speed of $v=\SI{10.9(5)}{km/s}$ implying a rotational period of about $P_{\rm rot}=\SI{7.7(2)}{days;}$  this is slightly higher than the orbital period of $P=\SI{6.0}{d}$.

The determination of the parameters of the stars is complicated by the grazing nature of the primary eclipse, creating a degeneracy between the inclination and the ratio of $(R_1+R_2)/a$ and $R_2/R_1,$ given the fact that the eclipse is basically described only by depth and duration. We tried to use the out-of-eclipse variability, which is evident in the light curve in Fig.~\ref{fig:1153lc}, to adress this issue. For a star with a temperature from $\SIrange{6700}{7000}{K}$ and $\log g=\SIrange{3.5}{4.5}{}$, the gravity-darkening coefficient should be around $\beta_1=\SI{0.26(2)}{}$ according to \citet{2003A&A...406..623C}. From the amplitude of the ellipsoidal variability $E=\SI{8.71(5)e-5}{}$ and the primary radius, we can derive the semi-major axis of the binary $a=\SI{18.3(3)}{R_\odot}$. Alternatively, we can use the mass estimate from isochrone fitting in {\sc Ariadne,} arriving at $a=\SI{18.35(11)}{R_\odot, which }$ is consistent with the previous value. We present our results under these assumptions separately in Table~\ref{tab:par}. 

Information about the secondary is very limited. We only obtained masses of $M_2=\SI{0.66(3)}{M_\odot}$ or $M_2=\SI{0.667(13)}{M_\odot,}$ with a reasonable precision level breaking the degeneracy using the ellipsoidal variation and isochrone fit for the primary, respectively. For the radius and temperature, we can only make crude estimations of $0.35<R_2<0.8$ from the shape of the primary eclipse and $T_2=\SI{3050(350)}{K}$ from the depth of the very shallow secondary eclipse. We note that the temperature estimate is highly uncertain - the value above does not consider possible changes in the secondary eclipse depth due to changes in eccentricity. For a star of the spectral class K6.5V, \citet{2013ApJS..208....9P} gave a mass of $M=\SI{0.66(2)}{M_\odot}$ and  radius of $R=\SI{0.65(1)}{R_\odot}$, which fall into our range and temperature of $T=\SI{4200(50)}{K,}$ which is significantly higher than our estimate.


\subsection{TOI-1615}
\begin{figure}[h!]
    \centering
    \includegraphics[width=\linewidth]{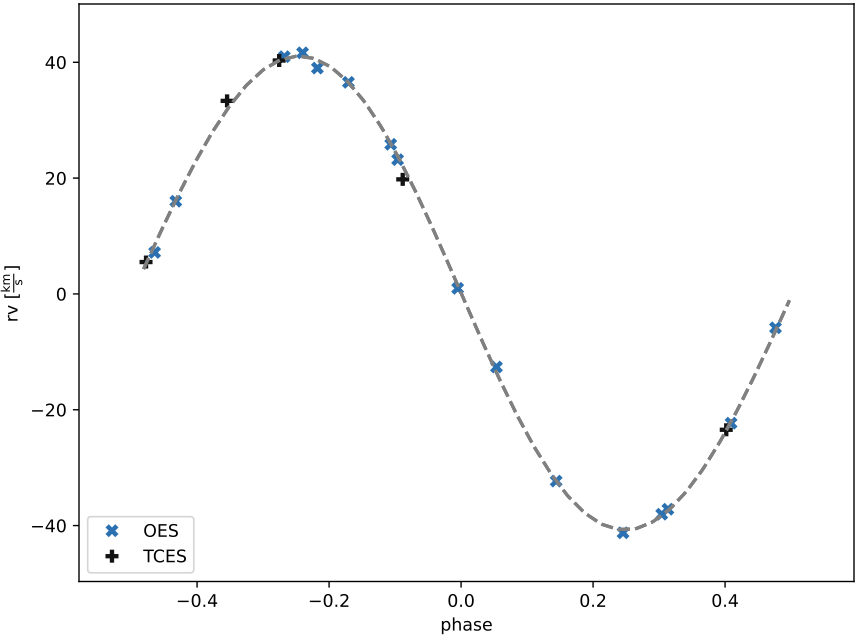}
    \caption{Radial velocity curve of TOI-1615 obtained by OES and TCES after correction for zero-point velocities with our best-fit model showing a circular orbit. Errors of the data points are smaller than the markers.}
    \label{fig:1615rv}
\end{figure}

The ten-magnitude star BD+76 916 (TIC 317507345) in the constellation of Cepheus was identified as an exoplanetary candidate, TOI-1615, by the TESS survey in December 2019. It was also recovered from the same data in an eclipsing binary study by \cite{2022ApJS..258...16P}. With a period of $\SI{3.94}{days}$, the light curve shows a V-shaped primary eclipse and a shallow secondary visible in the binned data from all available sectors. There is also a quasi-periodic variation with a frequency of about six cycles per day, which we attribute to $\delta$-Sct (DSCT) variability of the primary star. This variation is eliminated from the light curve by binning the photometric data phased with the orbital period. There is no imaging available in the ExoFOP database; however, there are a few faint stars in the surroundings with a TESS contamination ratio of about $0.014$. The star has a moderately high astrometric solution error of $\mathrm{RUWE}=1.446$, which is higher than what is typical for an isolated target, but still not so excessive that it can unequivocally indicate a companion. Thus, we consider the binary system as alone. We have 17 radial-velocity measurements from Ond\v{r}ejov at our disposal plotted in Fig.~\ref{fig:1615rv}.

The temperature of the primary star was estimated by {\sc Ariadne} as $T_1=\SI{7380(80)}{K}$ after the binary correction with a radius of $R_1=\SI{1.498(15)}{R_\odot}$. The spectroscopic work by A. Bieryla at TRES@FLWO in the ExoFOP archive shows the temperatures of $\SI{7212(73)}{K}$ and $\SI{6935(88)}{K}$ for the two spectra obtained, gravity of $\log g=4.2(1)$, metallicity of $[\mathrm{Fe/H}]=\SI{0.1(1)}{}$, and $v\sin i=\SI{12(1)}{km/s}$. In the case of parallel rotational and orbital axes, this yields a rotation period around $P_{\rm rot}=\SI{6.3}{d}$, which is clearly slower than the orbital period $P=\SI{3.94}{d}$.

\begin{figure}[h!]
    \centering
    \includegraphics[width=\linewidth]{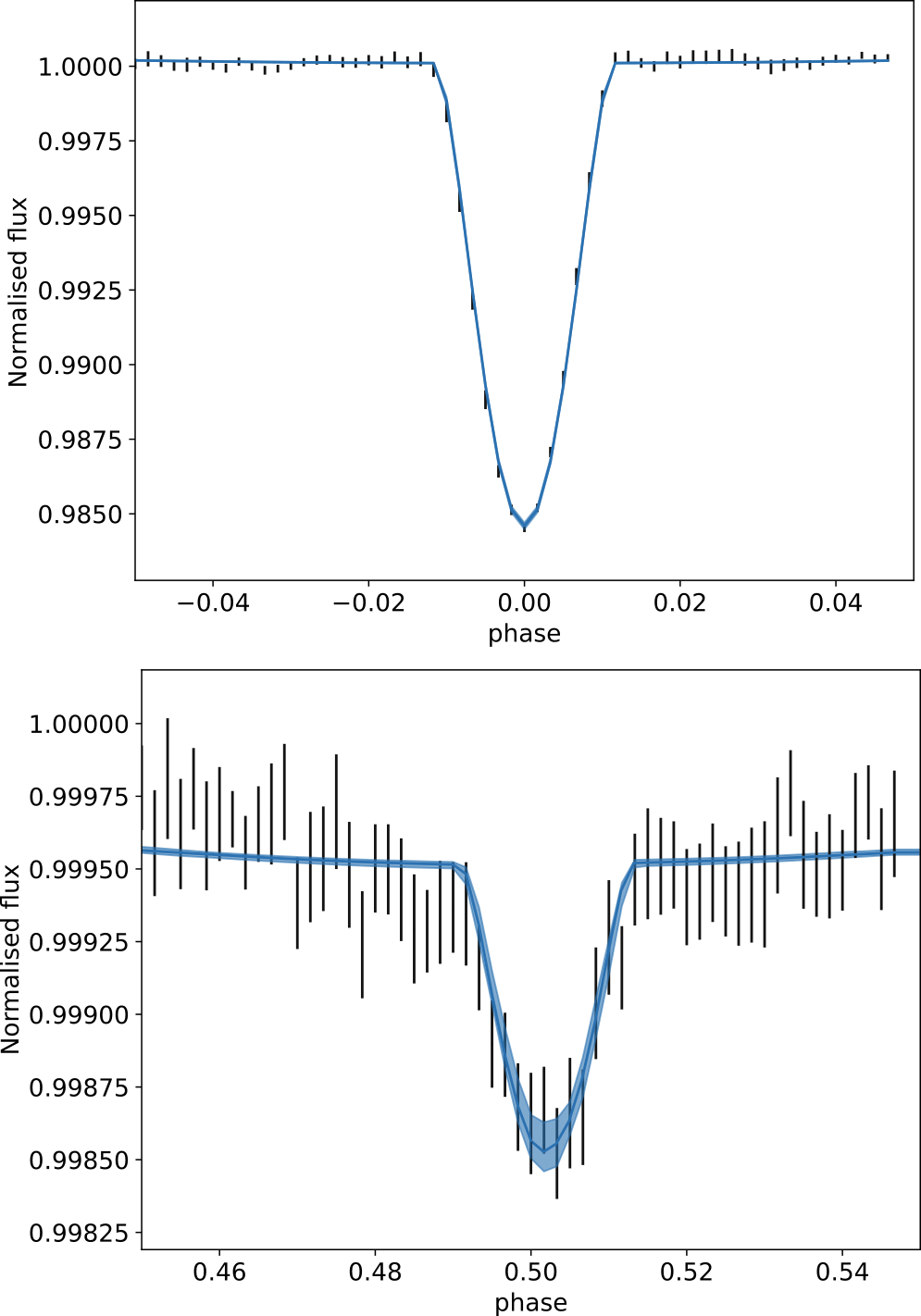}
    \caption{Light curve of of TOI-1615. We can see the partial primary eclipse in the top panel and very shallow secondary eclipse in the bottom panel.}
    \label{fig:1615lc}
\end{figure}

\begin{figure}[h!]
    \centering
    \includegraphics[width=\linewidth]{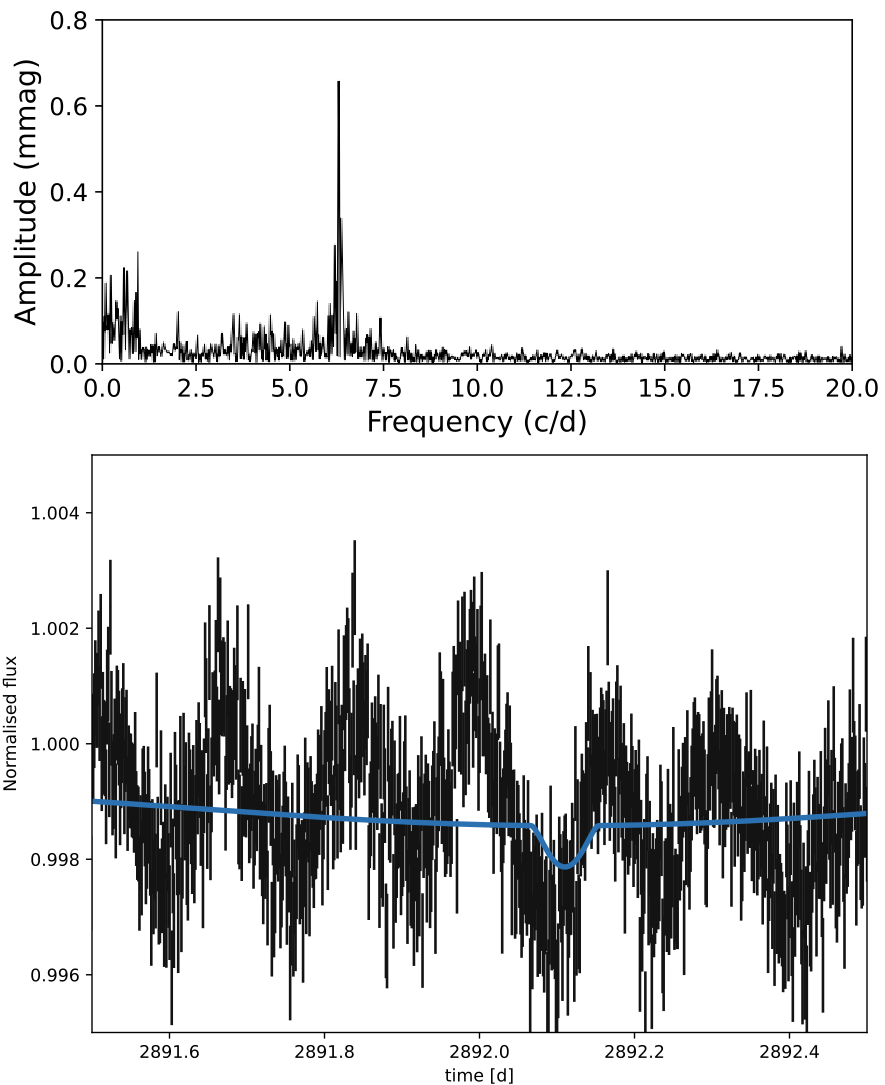}
    \caption{Delta Scuti pulsations of TOI-1615A. Top panel: Frequency-amplitude spectrum of the residual fluxes after removing the binary solution. Bottom panel: Zoomed-in view of pulsation in the normalized TESS light curve; blue curve corresponds to binary fit showing the depth of the secondary eclipse in comparison to the pulsation amplitude.}
    \label{fig:1615DSCT}
\end{figure}

Our combined RV+photometric model reveals an almost circular orbit with an orbital velocity semi-amplitude of $K_1=\SI{40.90(17)}{km/s}$. However, there is some degeneracy in the $i$, $R_2/R_1$, $(R_1+R_2)/a$ space due to the V-shaped primary and secondary eclipses presented in Fig.~\ref{fig:1615lc}. We thus provide two resulting sets of parameters in Table~\ref{tab:par} - with this degeneracy and the additional use of primary mass from isochrone fit in italics. We can see that the masses of the stars are only determined with 10\% accuracy without this aid. The primary star is of spectral class A9 judging by the temperature, but with a diameter of only 86\% of the expected value for a main-sequence star \citep{2013ApJS..208....9P}. Both of the mass estimates are in accordance with the hotter case. If we took the TRES temperatures and our radius, the star would still be too small in radius to be a main-sequence star. The secondary is an early M dwarf with estimates of M1, M2.5, and M3 from mass, temperature, and radius, respectively. This system is thus not described well as a pair of main-sequence stars. There are two possible explanations:
\begin{itemize}
\item our {\sc Ariadne} fit is "too hot"; a colder star could allow a larger primary radius in the SED and thus a bigger secondary.
\item the fit is right, but the stars are not evolving as well-separated main-sequence stars.
\end{itemize}
After the binary model removal from the light curve, we identified the main pulsation frequency of $f_0=\SI{6.303}{d^{-1}}$ and amplitude $0.00066$ relatively to the light of both stars, with S/N of 9.2. There are also secondary frequencies of $f_1=\SI{6.376}{d^{-1}}$ and $f_2=\SI{6.197}{d^{-1}}$ present with about half the amplitude of $f_0$. This variability presented in Fig. \ref{fig:1615DSCT} combined with the derived effective temperature clearly shows that the primary of TOI-1615 is a multi-mode DSCT. 

It is worth noting that we can neglect the effect of pulsations on the radial-velocity values and the estimation of stellar parameters from the binary model. The ratio of radial velocities and visual amplitudes of DSCT stars is between 50 and 125\,km$^{-1}$mag$^{-1}$ \citep{1976ApJ...210..163B}. The amplitude of pulsations is about 3\,mmag (see Fig.~\ref{fig:1615DSCT}), which translates to an expected amplitude of 0.15\,$-$\,0.375\,km/s. In addition, the exposure time of the individual exposures was one hour (1/4 of the pulsation cycle), which decreases the observable radial-velocity amplitude of the pulsations. These values are comparable to the uncertainties of our radial-velocity observations (Table~\ref{Tab:RVs}). Finally, the observations were taken randomly with respect to the pulsation cycle, so the effect of pulsations averages out.

\subsection{TOI-1788}
\begin{figure}[h!]
    \centering
    \includegraphics[width=\linewidth]{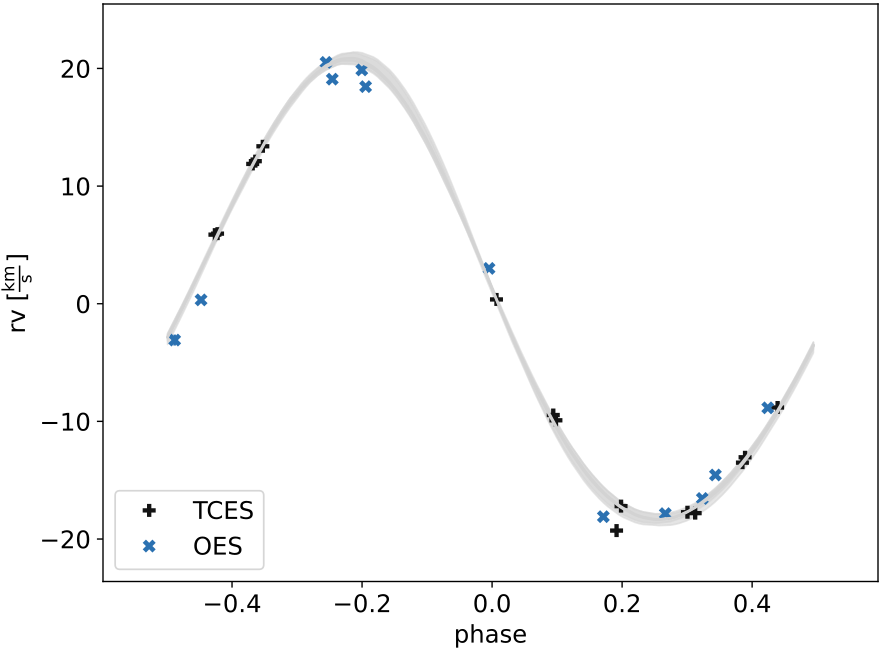}
    \caption{Radial-velocity curve of TOI-1788 obtained by OES and TCES after correction for instrumental zero-point shift. The shaded region demonstrates uncertainty in model posteriors, and errors of measured data points are smaller than the symbols. The higher than expected scatter is caused by stellar spots on the primary.}
    \label{fig:1788rv}
\end{figure}

In Leo Minor, there is an eleventh magnitude star TYC 2518-812-1 (TIC 450327768) that was identified as TESS exoplanetary candidate TOI-1788 in March 2020. It was also identified as a spectroscopic binary by \cite{2023A&A...674A..34G}. There is no companion present in an imaging study using NESSI@WIYN by Mark Everett (with a sensitivity of 3.5$^m$ down to $0.15''$). It has a renormalized astrometric excess error of $\mathrm{RUWE}=1.028$ in Gaia and a negligible TESS contamination ratio, suggesting a lone system. The TESS light curve shows four primary eclipses with a period of $P=\SI{5.344}{d}$ and no secondary eclipses visible; there are also sinusoidal changes in total light, with a period of $P=\SI{5.549}{d}$ and an amplitude up to $0.02$ in normalized flux in the only available short cadence sector from February 2022.  We detrended this variability prior to the binary parameter fit as described in Sect. \ref{sec:tess}. We discuss its nature later. We have at our disposal 13 radial-velocity measurements from the Ond\v{r}ejov Observatory and 18 from Tautenburg Observatory, which are displayed in Fig.~\ref{fig:1788rv}. There is also one long cadence {\sc TESS} sector (Sect. 21) with four additional eclipses available; however, it contains only three or four points per eclipse. Also notable is the partial observation of the eclipse from the ground on January 21, 2022 UT with the MuSCAT2 multicolor imager \citep{2019JATIS...5a5001N} installed on the 1.5~m Telescopio Carlos S\'anchez (TCS) at the Teide Observatory, Spain, with the eclipse center at $T_c= \SI{2459601.7796(9)}{}$. We only used these to improve the linear ephemeris of the system. 

The primary star has, according to the {\sc Ariadne} fit, a temperature of $T_1=\SI{5760(30)}{K}$ and radius $R_1=\SI{1.26(2)}{R_\odot}$. The spectral fit by A. Bieryla at FLWO@TRES available on the ExoFOP database gives $T_1=\SI{5720(50)}{K}$, $\log g=4.39(10)$, $[\mathrm{Fe/H}]=0.00(8)$ and a rotational velocity of $v \sin i=\SI{12.2(5)}{km/s}$ with a corresponding rotational period of $P_{\rm rot}=\SI{5.24(23)}{d,}$ which is similar to both the orbital period and the period of photometric variability.

\begin{figure}[h!]
    \centering
    \includegraphics[width=\linewidth]{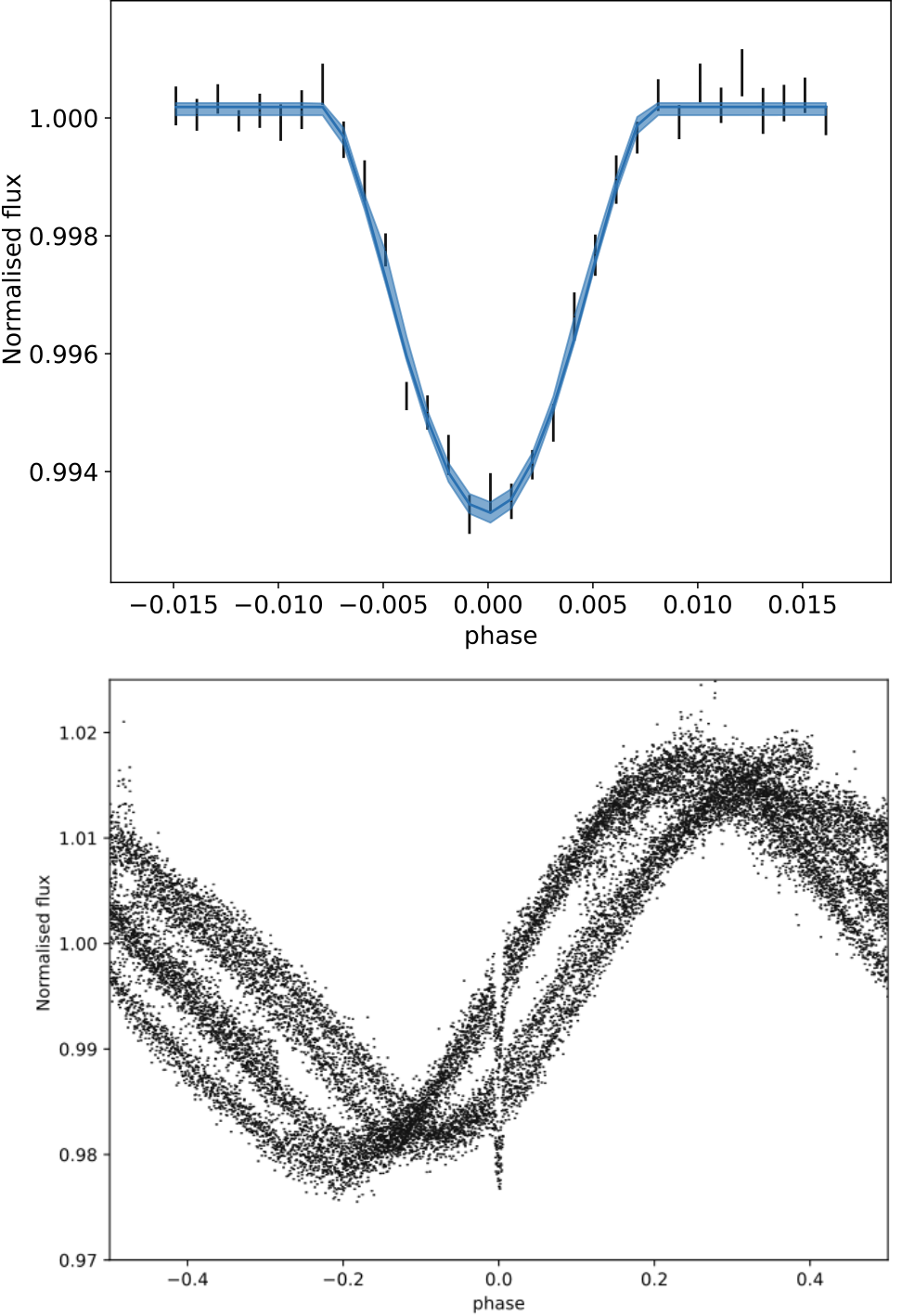}
    \caption{Light curve of TOI-1788. In the top panel, we can see the primary eclipse. We do not detect any secondary eclipses. Rotation-related changes in the original TESS-SPOC light curve shown in the orbital phase are located in the bottom panel.}
    \label{fig:1788lc}
\end{figure}

In Fig.~\ref{fig:1788lc}, we can see the resulting fit of the binned light curve. The fit is degenerate in $i$, $R_2/R_1$ and $(R_1+R_2)/a$ with only a lower bound on the secondary radius $R_2>\SI{0.09}{R_\odot}$ and a crude estimate of semi-major axis $a=\SI{13.7(20)}{R_\odot}$. As we did not detect the secondary eclipse, it is best to consider the system just as a spectroscopic binary with known inclination $i=\SI{84(2)}{\degree}$. If we use the {\sc Ariadne}-MIST model, we obtain masses for the components of $M_1=\SI{0.96(7)}{M_\odot}$ -- a G2IV star; and $M_2=\SI{0.178(15)}{M_\odot}$ -- most probably an M4V star. The full set of estimated parameters can be found in Table~\ref{tab:par}.

\begin{table*}[h!]
\caption{Derived parameters of the orbit and component stars from SED, RV, and photometry, with the additional use of MIST isochrone fit for the primary in italics.}
\centering
\begin{tabular}{c|c c c c c}
Target & TOI-416&TOI-1143&TOI-1153&TOI-1615&TOI-1788\\ 
\hline
orbital period $P$ [d] &\SI{7.011396(3)}{}&\SI{11.207189(17)}{}&\SI{6.036078(3)}{}&\SI{3.9401793(6)}{}&\SI{5.343604(6)}{}\\
superior conjunction $t_0$ [d] &\multirow{2}{*}{\SI{1413.19585(17)}{}}&\multirow{2}{*}{\SI{2400.1247(6)}{}}&\multirow{2}{*}{\SI{1685.5773(3)}{}}&\multirow{2}{*}{\SI{1790.82286(13)}{}}&\multirow{2}{*}{\SI{2612.4629(8)}{}}\\
\small{(BJD - \num{2457000})} &  &  &  &  &  \\ 
semi-major axis $a$ [$\mathrm{R_\odot}$] &\SI{17.6(3)}{}&\SI{24.8(3)}{}&\SI{18.3(3)}{}/\textit{18.35(11)}{}&\SI{14.1(4)}{}/\textit{13.44(9)}&\SI{13.7(20)}{}/\textit{13.4(5)}\\
eccentricity $e$  &$<\SI{0.010}{}$&\SI{0.285(4)}{}&\SI{0.0032(11)}{}&\SI{0.011(4)}{}&\SI{0.070(7)}{}\\
argument of periastron $\omega$ [$\si{\degree}$]        &--&\SI{238.0(5)}{}&\SI{13(31)}{}&\SI{68(12)}{}&\SI{33(11)}{}\\
inclination $i$ [$\si{\degree}$] & \SI{86.12(13)}{}&\SI{86.24(5)}{}&\SI{83.1(7)}{}&\SI{83.6(5)}{}&\SI{84(2)}{}\\
\hline\noalign{\smallskip}
RV semi-amplitude $K_1$ [km/s]  &\SI{11.327(8)}{}&\SI{8.44(7)}{}&\SI{44.74(8)}{}&\SI{40.90(17)}{}&\SI{19.6(4)}{}\\
mass function $f(M)$ [$\mathrm{M_\odot}$]               &\SI{0.001056(3)}{}&\SI{0.000 623(16)}{}&\SI{0.0560(3)}{}&\SI{0.02792(34)}{}&\SI{0.00412(24)}{}\\
mass ratio $q$                  &\SI{0.0976(22)}{}&\SI{0.0783(12)}{}&\SI{0.415(6)}{}/\textit{0.415(3)}&\SI{0.294(13)}{}/\textit{0.313(4)}&\SI{0.18(3)}{}\textit{0.183(8)}\\
TESS third light $l_3$ [\%] &--&\SI{12.9(10)}{}&--&--&--\\
\hline
\multicolumn{6}{c}{primary component}\\
\hline
volumetric radius $R_1$ $[\mathrm{R_\odot}]$    & \SI{1.82(2)}{}        &\SI{2.07(3)}{}&\SI{1.67(2)}{}&\SI{1.498(15)}{}&\SI{1.26(2)}{}\\
mass $M_1$ $[\mathrm{M_\odot}]$ & \SI{1.36(7)}{}        &\SI{1.51(7)}{}& \SI{1.60(6)}{}/\textit{1.61(3)} &\SI{1.87(16)}{}/\textit{1.60(3)}&\SI{1.1(5)}{}/\textit{0.96(7)}  \\
effective temperature $T_1$ [K] & \SI{6330(60)}{}       &\SI{6300(50)}{}& \SI{7000(130)}{}&\SI{7380(80)}{}&\SI{5760(30)}{}\\
surface gravity $\log_{10} [g_1]_{cgs}$ & \SI{4.05(3)}{}        & \SI{3.99(2)}{}& 4.20(2)/\textit{4.20(1)}& 4.36(4)/\textit{4.29(1)} & 4.3(3)/\textit{4.22(4)}\\
mean density $\log_{10} [\rho_1]_{cgs}$ & \SI{-0.49(3)}{}       &\SI{-0.62(2)}{}&-0.31(3)/\textit{-0.31(2)}&-0.11(5)/\textit{-0.17(2)}&-0.1(3)/\textit{-0.17(4)}\\
\hline
\multicolumn{6}{c}{secondary component}\\
\hline
volumetric radius $R_2$ $[\mathrm{R_\odot}]$    &\SI{0.157(2)}{}&\SI{0.142(3)}{}&$0.35<x<0.8$&\SI{0.32(7)}{}&$>0.09$\\
mass $M_2$      $[\mathrm{M_\odot}]$    &\SI{0.132(8)}{}&\SI{0.118(5)}{}&\SI{0.66(3)}{}/\textit{0.667(13)}&\SI{0.55(5)}{}/\textit{0.501(14)}&\SI{0.19(10)}{}/\textit{0.178(15)}\\
effective temperature $T_2$ [K] &\SI{3010(40)}{}&\SI{3210(60)}{}&\SI{3350(550)}{}&\SI{3480(80)}{}&--\\
surface gravity $\log_{10} [g_2]_{cgs}$ &\SI{5.17(3)}{}&\SI{5.21(3)}{}&--&5.2(2)/\textit{5.1(2)}&--\\
mean density $\log_{10} [\rho_2]_{cgs}$ &\SI{1.68(3)}{}&\SI{1.76(3)}{}&--&\SI{1.3(4)}{}/\textit{1.3(4)}&--\\
\end{tabular}
\label{tab:par}
\end{table*}

Next, we return to the nature of the photometric variability. The variability certainly does not originate from the secondary, given that the amplitude of 2\% is much higher than the light contribution of the secondary itself. The variability can be detected in broadband (with a sensitivity level in range of $\SIrange{400}{700}{nm}$) SuperWASP photometry data \citep{2006PASP..118.1407P} from December 2006 - May 2007 with an amplitude of $0.0125^{\rm m}$ and a period of $P_{SW}=\SI{5.62}{d,}$ as well as with half this period and an amplitude of $0.0066^{\rm m}$. The period is slightly longer than the one in the TESS data; the latter was, however, determined from just one month-long series of data in Sector 48. The amplitude is slightly lower in SuperWASP data with a more prominent harmonic component. In Sector 21, the variability has a more pronounced double peak nature evolving across the sector.
We considered the following options to explain this:
\begin{itemize}
\item pulsations of the primary;
\item starspots on the primary;
\item a blend with a background variable.
\end{itemize}
We did not notice another set of spectral lines in the TOI-1788 spectra, nor did we notice a secondary bump in the cross-correlation function. This, together with the imaging, rules out a background variable that could contribute enough light to explain the variability. 
Pulsations of the primary are not expected. In this temperature range, one would only expect faster, solar-type oscillations with periods up to a few minutes and low amplitude. Star spots are thus the most probable source of this variability, which is further supported by the presence of the double harmonic frequency. 
The difference between the photometric periods and the orbital one can be explained by differential rotation of a spin-synchronized primary with a large spotted region near one of the star poles. We tried to check for changes in depth or duration of the primary eclipses, but given their shallow nature we were not able to detect any changes conclusively. We also note that the stellar spots are the most probable source of higher scatter in radial velocities in Fig.~\ref{fig:1788rv}, just as the corresponding errors of the data points suggest.

Finally, the binary has a possible outer companion with the designation 2MASS 10403438+3543348 at a distance of approximately $\SI{13}{",}$  a projected relative velocity of about $\mu=\SI{0.49}{mas/yr,}$ and a reported difference in parallax of only $\SI{0.14(11)}{mas}$. At the distance of the sources, the projected separation is $d\approx\SI{3000}{au}$ and the projected relative velocity is  $v\approx\SI{0.54}{km/s,}$ which is consistent with a companion on an orbit of a few hundred thousand years. 
Both the {\sc Gaia} color index  $\mathrm{BP}-\mathrm{RP}=2.46$ and apparent magnitude $\mathrm{G}=16.688$ suggest an M3 main-sequence dwarf star. 
\section{Discussion}
\label{sec:discussion}
We present parameters for five eclipsing binaries that are also single-lined spectroscopic binaries. To obtain the primary star parameters, we used SED modeling; the radii and temperatures obtained thus depend on the precision of the Gaia DR2 parallaxes used and the contamination of the photometry by other sources. Our stars are fainter than $\mathrm{G}<8$ and lie in relatively sparse fields, thus avoiding the known systematic issues outlined on the Gaia website, \footnote{\url{https://www.cosmos.esa.int/web/gaia/dr2-known-issues\#AstrometryConsiderations}} but we did not investigate the other issues raised in \cite{2018A&A...616A...2L}. 

\begin{figure*}[h!]
\centering
\includegraphics[width=\linewidth]{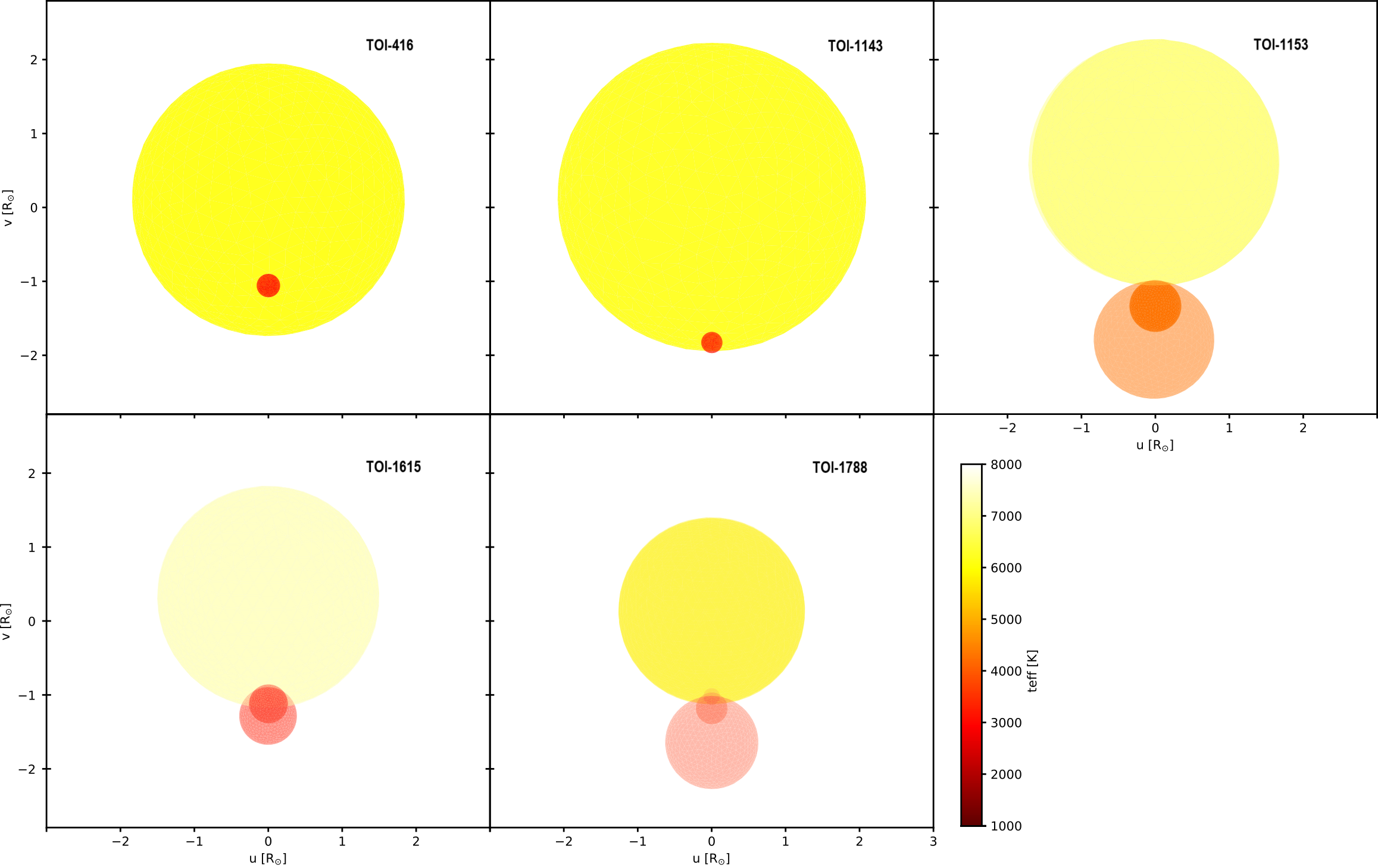}
\caption[]{Family portraits of the systems in the middle of the primary eclipse. The~secondary moves horizontally across the disk of the primary. Shading for the secondary corresponds to the uncertainty in the posterior values of secondary radius. All systems are drawn to the same scale within  boxes that are $\SI{6}{R_\odot}$ in length. 
}
\label{fig:family}
\end{figure*}

We fit the extinction, which is rather small for all systems. To rule out contamination by other point sources close to the targets, we used the astrometric solution excess error RUWE in Gaia and speckle imaging in the Exofop archive, where available. A high RUWE value of $>1.4$ is a good indicator of binarity; however, a smaller value does not rule out possible companion(s). \citet{2024A&A...682A...7B} showed that $\approx 40\%$ of binaries in the ninth catalog of spectroscopic binary orbits \citep{2004A&A...424..727P} have values of $\mathrm{RUWE}<1.4$, mostly in cases of targets with shorter orbital periods below $\SI{100}{d}$. For TOI-1615, there was no further imaging available. In the other cases, we found reasonable contrast down to $0.1"-0.2"$, which at the distance of the targets gives a region of radius $\SIrange{15}{30}{au}$. Another bound star could still orbit the systems on an orbit up to $\approx \SI{100}{y}$. If we consider a mass ratio of $q=0.3$, the third component would cause radial-velocity variability in the central binary system with a semi-amplitude of about  $\SIrange{2}{3}{km/s,}$ which can easily be missed in our data with time spans of a few years as well as in the Gaia astrometry. Also, the change in times of eclipses is quite small in this case -- just a five minute or less change in one year of observations. Thus, it would be preferable to place tighter constraints on the possible companions with more sensitive imaging or long-term eclipse time monitoring in order to be certain for the well-described systems TOI-416 and TOI-1143. Another issue for TOI-1153, TOI-1615, and TOI-1788 is that the lack of knowledge of the secondary parameters hinders us in the removal of its contribution to the SED. Here, we used the expected values for stars of the same mass from \cite{2013ApJS..208....9P} instead of our own radius and temperature estimates with large uncertainties.

Another important point is the use of tabulated values of limb-darkening coefficients, albedos, gravity brightening coefficients, and beaming coefficients for the photometry model. As can be seen in the works of \cite{2017A&A...600A..30C}, \cite{2001MNRAS.327..989C}, \cite{2003A&A...406..623C}, and \cite{2020A&A...641A.157C}, respectively, limb darkening is a well-studied effect with observational verification of the models compared to the other coefficients, where precise observations are in short supply. Limb darkening manifests in the shape of the eclipse intertwined with the eclipse geometry, so we opted to use the tables. In the case of albedos and gravity brightening, there is also an abrupt change of the values around the transition from a convective envelope to a radiative one in the $T=\SIrange{6000}{7000}{K}$ range, which is relevant for four of our targets. We used fixed values of these parameters as outlined in Sect. \ref{sec:Phoebe}. This could cause model-dependent errors with secondary eclipse depth, secondary temperature being the most vulnerable. We discuss the beaming coefficients in a later section.
\FloatBarrier

\subsection{Radius-mass-temperature relationships}
\begin{figure}[h!]
    \centering
    \includegraphics[width=\linewidth]{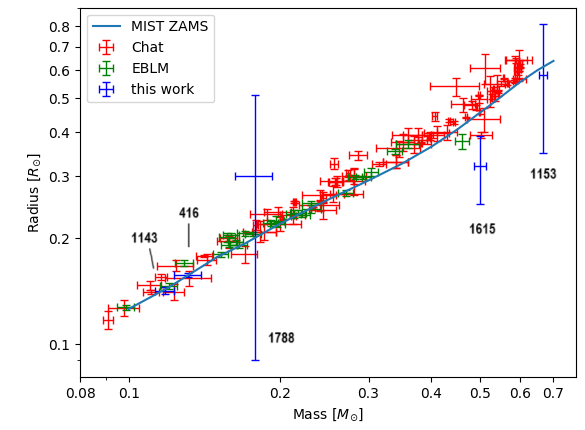}
    \caption{Radius-mass diagram of dwarf stars. Stars from \cite{2018AJ....156...27C} and references therein are shown in red, stars from the EBLM project \cite{2023arXiv231211339S} in green, and our secondaries in blue. The blue line denotes the zero-age main sequence from the MIST stellar models for the metallicity [Fe/H]=0.0.}
    \label{fig:radius-mass}
    \footnotesize 
\end{figure}

One of the most important uses of our work is the calibration of fundamental stellar parameters for dwarf stars. In Fig. \ref{fig:radius-mass}, we can see the radius-mass relation for the secondaries from our systems together with dwarf stars parameters from \citet[][ references therein,]{2018AJ....156...27C} and \cite{2023arXiv231211339S}, which contains the work of the EBLM project. We can see that for stars with radiative interiors with masses above $\SI{0.35}{M_\odot}$, the observed radii are inflated in comparison with the MIST stellar models' zero-age main sequence for a metallicity of [Fe/H]=0.0. For the fully convective stars,  $M<\SI{0.35}{M_\odot}$, the model describes the measurements well. Our points with robust measurements for TOI-416 and TOI-1143 lie amongst the other known stars in the R-M diagram close to the theoretical prediction. 

\begin{figure}[h!]
    \centering
    \includegraphics[width=\linewidth]{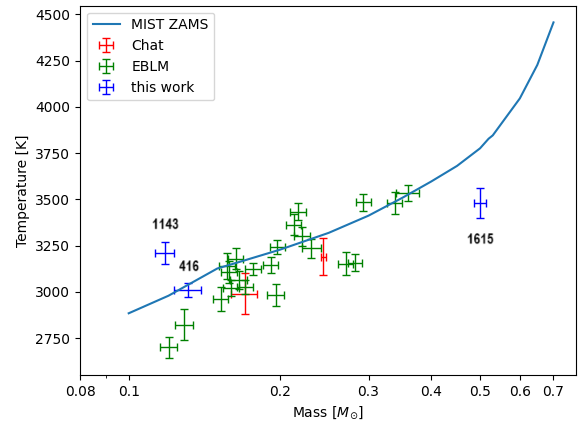}
    \caption{Temperature-mass diagram of dwarf stars. Stars from \cite{2018AJ....156...27C} and references therein are shown in red, stars from the EBLM project \cite{2023arXiv231211339S} in green, and our secondaries with temperature estimates in blue. The blue line denotes the zero-age main sequence from the MIST stellar models for metallicity [Fe/H]=0.0.}
    \label{fig:teff-mass}
    \footnotesize 
\end{figure}

The temperature-mass diagram in Fig. \ref{fig:teff-mass} shows a scatter with most of the points colder than the model, except for the TOI-1143, which is clearly above. However, considering the error bars, the measured and model values agree within a few hundred kelvin. We can also use these relations to predict the expected radii and temperatures for the other three secondaries - $R_2=\SI{0.65(4)}{R_\odot}$, $T_2=\SI{4300(200)}{K}$ for TOI-1153B; $R_2=\SI{0.50(2)}{R_\odot}$, $T_2=\SI{3750(150)}{K}$ for TOI-1615B; and $R_2=\SI{0.18(2)}{R_\odot}$, $T_2=\SI{3150(150)}{K}$ for TOI-1788B.

\subsection{Beaming and boosting}
As described in Table \ref{tab:beam-fit}, we have three systems with beaming amplitudes estimated: TOI-416, TOI-1153,  and TOI-1615. We can compare these with the theoretical predictions and the relation and values according to \cite{2011MNRAS.415.3921F} and \cite{2020A&A...641A.157C}. Our parameter $B$ is related to the beaming coefficients in TESS band $B_T$ in tables by \cite{2020A&A...641A.157C} in the following way:

\begin{equation}
    B=l_1 B_{T,1}\frac{K_1}{c}-l_2 B_{T,2}\frac{K_2}{c}=\left(l_1 B_{T,1}-\frac{(1-l_1) B_{T,2}}{q}\right)\frac{K_1}{c}\,,
\end{equation}

\noindent where $l_i$ is the relative light contribution of the component in the pass band, $K_i$ is the semi-amplitude of the radial velocities of this component, and $q$ is the binary-mass ratio. Here, it is important to note that the contribution of the second term tend to be negligible as the mass ratio becomes smaller; for FGKM stars, it approximately holds $R\propto M$, and so $l_2/l_1$ decreases faster than $q^2$ given by the additional decrease also in temperature. We can see a comparison between the theoretically expected values, $B_{exp}$, and the measured ones, $B_{meas}$, in Table \ref{Tab:beaming}. In the case of the well-characterized TOI-416, the values agree nicely. For TOI-1153 and TOI-1615, the measured values are significantly lower, pointing to the issues with parameter estimation. This discrepancy can be explained only if we underestimated the temperature and radius values of the secondaries in these systems. Reintroducing support for beaming into a future version of PHOEBE is planned (DJ, private communication), requiring the user to manually provide the beaming coefficients. This could be used as an additional constraint to break the degeneracy in grazing eclipses, but the beaming coefficients strongly depend on temperature, which is not easy to determine for the secondary.

\begin{table}
\caption{Beaming amplitudes for TOI-416, TOI-1153, and TOI-1615.}
\label{Tab:beaming}
\centering
\begin{tabular}{c|ccc}
Target & TOI-416 & TOI-1153& TOI-1615\\
\hline
$B_{T,1}$ & 3.05(3)&2.75(5)&2.64(3)\\
$B_{T,2}$ & 6.2(3)&6.2(5)&5.6(1)\\
$l_1$& 0.99978(3)& >0.990 & 0.998(1) \\
$B_{exp}\cdot 10^5$&11.3(2)&39.5(15)&35.5(6)\\
$B_{meas}\cdot 10^5$&10.9(14)&24.3(25)&25.6(92)\\
\end{tabular}
\end{table}


\subsection{Grazing eclipses}
As we can see from Fig.~\ref{fig:family}, grazing eclipses lead to degeneracy in the stellar radii and inclination, giving the same photometric profile during the eclipse for vastly different radii of the stars. We now look more closely at the limiting case of just grazing eclipses.  We assume a circular orbit for the demonstration. The duration of the eclipse ($T$) can be derived from the orbital speed ($v$) and the distance between centers of the secondary during first and last contact ($2L$). For the length ($L$), we have the following from the right triangle:

\begin{equation}
    L^2+a^2\cos^2i=(R_1+R_2)^2\,.
\end{equation}
If we use an equation for the orbital speed $v=2\pi a/P,$ we obtain a relation between the directly measurable quantity $T/P$ and the derived ones:
\begin{equation}
\label{eq:duration}
    \left(\frac{\pi T}{P}\right)^2 + \cos^2 i = \left(\frac{R_1+R_2}{a}\right)^2\,.
\end{equation}
Similarly, for the separation of stellar disk centers ($H(t)$) in time measured from the middle of the eclipse, we can obtain an equation:
\begin{equation}
\label{eq:H}
    H^2(t)=(R_1+R_2)^2+\left(2\pi a\right)^2\left[ \left(\frac{t}{P}\right)^2-\left( \frac{T}{2P}\right)^2 \right]\,.
\end{equation}

The depth of the eclipse ($\delta$) under the assumption of homogeneous disks can be expressed as
\begin{equation}
\label{eq:depth}
    \delta=\frac{A}{\pi R_1^2+\pi\alpha R_2^2}\,,
\end{equation}
where $A$ is the covered area of the primary and $\alpha$ is the surface-brightness ratio of the secondary to the primary. The area of the thin lens in the case of a grazing eclipse can be expressed as
\begin{equation}
\label{eq:area}
    A=\frac{2}{3}\theta_1^3 R_1^2+ \frac{2}{3} \theta_2^3R_2^2
,\end{equation}
with an error term in $\theta_i^5$, where $\theta_i$ are the angles between the end of the lens and its center as seen from the center of the respective stellar disks. From the common chord, we have the equation $R_1\sin \theta_1=R_2 \sin \theta_2$, and from the segment between the stellar centers we have $R_1\cos\theta_1+R_2\cos\theta_2=H(t=0)$. Using the Taylor series, we obtain
\begin{equation}
    R_1\theta_1=R_2\theta_2\,,\qquad R_1 \frac{\theta_1^2}{2}+R_2 \frac{\theta_2^2}{2}=R_1+R_2-H(t=0)\,.
\end{equation}
If we define $E=R_1+R_2-H$ (which is our limiting quantity), we obtain the angles
\begin{equation}
    \theta_1=\sqrt{\frac{2R_2 E}{R_1(R_1+R_2)}}\,,\qquad \theta_2=\sqrt{\frac{2R_1 E}{R_2(R_1+R_2)}}\,.
\end{equation}
We can see that for $R_2<R_1$ is $\theta_2>\theta_1,$ and we need $E\ll R_2<R_1$ for the approximation to hold. Substituting into Equations~\ref{eq:area} and~\ref{eq:depth}, we obtain
\begin{equation}
    \delta=\frac{2}{3\pi}\left(\frac{2E}{R_1+R_2}\right)^\frac{3}{2}\frac{\sqrt{R_1R_2}(R_1+R_2)}{R_1^2+\alpha R_2^2}
.\end{equation}

From Equation~\ref{eq:H}, we can derive the approximate relation
\begin{equation}
    E=\left(2\pi a \right)^2\frac{1}{2(R_1+R_2)}\left[ \left( \frac{T}{2P}\right)^2 -\left(\frac{t}{P}\right)^2\right]\,.
\end{equation}

If we substitute this into the previous equation, we obtain the time dependence of the depth
\begin{equation}
\label{eq:depth2}
    \delta(t)=\frac{2}{3\pi}\left(\frac{2\pi a}{R_1+R_2}\right)^3\frac{\sqrt{R_1R_2}(R_1+R_2)}{R_1^2+\alpha R_2^2}\left[ \left( \frac{T}{2P}\right)^2 -\left(\frac{t}{P}\right)^2\right]^\frac{3}{2}\,.
\end{equation}
We thus obtain Equations~\ref{eq:depth2} and~\ref{eq:duration} for the eclipse depth as a function of time $t$ ($\left| t \right|<T/2$) and its duration: equations relating two measurable quantities and the three unknown parameters $i$, $R_2/R_1$ and $(R_1+R_2)/a$.

An easy way to check for the presence of degeneracy in a shallow eclipse is to estimate the fractional width of the eclipse profile at half the depth $\xi,$  that is, the ratio of the full width at half minima (FWHM) to the duration $T$ between first and last contact. $\xi$ should take values from close to $\xi\approx 1.0$ in the case of total eclipses (in practice slightly lower due to limb darkening) with trapezoidal shapes, to $\xi\doteq 0.608$ in the degenerate case with a more V-shaped eclipse. 

\begin{figure}[h]
    \centering
    \includegraphics[width=\linewidth]{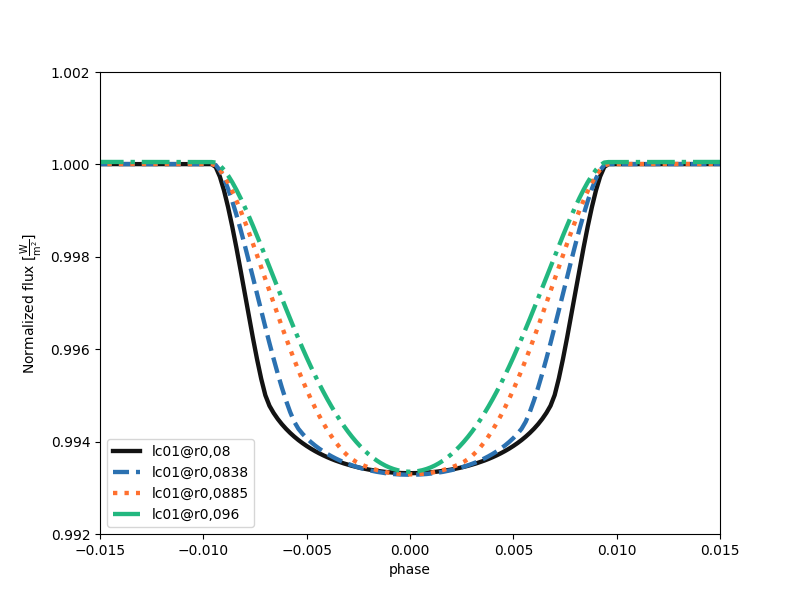}
    \caption{Light curve of a primary eclipse with fixed depth and duration. We demonstrate the changing shape during transition from total eclipse to the degenerate grazing case. From the outside curve inward, we used values $R_2/R_1=[0.0800,0.0838,0.0885,0.0960]$.  For $R_2/R_1>0.1$, the change of shape is indistinguishable from the $R_2/R_1=0.0960$ case.}
    \label{fig:tangent}
\end{figure}

As an example, we consider a system with an eclipse of maximal depth of $\delta_{\rm max}=0.0064$ and duration of $T=0.02 P$. Using {\sc Phoebe}, we took a primary with a radius of $R_1=\SI{1.67}{ R_\odot}$ and a temperature of $T_1=\SI{7000}{K}$ akin to the F1V star and tried to obtain the eclipse depth and duration for a sample of main-sequence secondaries (adjusting the temperature for the given radius) on $P=\SI{10}{d}$ orbit. We first fixed the inclination and then adjusted $(R_1+R_2)/a$ to fit the duration and, subsequently, $R_2/R_1$ to fit the depth. The resulting light curve in the Johnson $V$ band in Fig.~\ref{fig:tangent} shows a quick change from the trapezoidal shape for the minimal possible value of $R_2/R_1=0.08$ at $i=\SI{87.05}{\degree}$ and $(R_1+R_2)/a=0.0789$ to the degenerate shape at $R_2/R_1=0.096$, $i=\SI{84.55}{\degree}$, $(R_1+R_2)/a=0.112$. For bigger radii of the secondary, the shape of the primary eclipse was practically indistinguishable all the way up to $R_2=R_1$ at $i=\SI{80.8}{\degree}$ and $(R_1+R_2)/a=0.1708$. 

The visible difference between the light curves was in the presence of a deepening secondary eclipse with increasing temperature of the secondary. This difference, however, can be lost in the case of an even slightly eccentric orbit. Therefore, the only usable difference is the presence of ellipsoidal variations that strengthen as the stars become closer in size. We demonstrated the use of ellipsoidal variation in breaking the degeneracy in the case of TOI-1153. In the case of TOI-1788, there was no helping factor given the spot-associated variability of the star. Another option to break the degeneracy is to have either an estimate for the secondary radius from the spectral-energy-distribution fit - feasible mostly when the secondary is an evolved giant star - or to have an SB2 binary, thus obtaining a semi-major axis from the radial velocities of both components and solving for $i$ and $R_2$ from the depth and duration. However, it is still necessary to use the spectral-energy-distribution fit to obtain the primary radius. One may also wonder whether it is possible to break the degeneracy using the information about stellar rotation from spectra. We can quickly dismiss this option with an example. If we consider the difference between $\SI{84}{\degree}$ and $\SI{85}{\degree}$ inclinations in our systems, the difference between the sines of these values is just $1.7\cdot 10^{-3}$. Even under the assumption that the orbital and rotational axis of the primary are colinear on a $<\SI{1}{\degree}$ scale (which we deem dubious), we would need the orbital period, projected rotational velocity, and primary radius all measured with a relative precision on the order of $10^{-3}$ or better, which is unfeasible.

\section{Conclusions}
\label{sec:concl}

We obtained the orbital elements and physical parameters of five eclipsing binary stars identified by the TESS mission and observed for radial-velocity follow-up from Ond\v{r}ejov Observatory, Tautenburg Observatory and others. Two of the systems are well described: TOI-416, consisting of an F6 star leaving the main sequence and an M5V secondary on a circular orbit; and TOI-1143, consisting of an F6 sub-giant and an M5.5V dwarf on an eccentric orbit with an outer companion. The parameters of these systems, with a precision level of about 2\% in radii and 5\% in masses, can be used in the calibration of the stellar models of dwarf stars. 

TOI-1153 ($\approx$F2+K6), with large ellipsoidal variation; TOI 1615 ($\approx$A9+M3), with a pulsating primary; and TOI-1778 ($\approx$G2+M5), with a suspected outer M3 companion are somewhat lacking in the full description given the poor constraints on the photometry-related parameters due to grazing eclipses.
We conclude that exoplanetary candidates are a rich source of eclipsing binaries of two types: fully eclipsing, low-mass stars in front of mid- to early-type primaries, allowing for a good description of the system; and grazing systems with larger mass ratios that provide only poor solutions for the photometric quantities, and thus also system parameters without further model-dependent assumptions.

\medskip
\begin{acknowledgements}
JL, PK, RK and MV would like to acknowledge the support of the GACR grant 22-30516K. JL would like to thank grant ANID-23-05 for covering the travel expenses. MS acknowledges the financial support of the Inter-transfer grant no LTT-20015. EG acknowledges the generous support by the Th\"uringer Ministerium f\"ur Wirtschaft, Wissenschaft und Digitale Gesellschaft. JL would like to acknowledge the funding of the stay in Chile, enabling the completing of the paper, from the grant EXOWORLD under HORIZON programme with a project number 101086149. DJ acknowledges support from the Agencia Estatal de Investigaci\'on del Ministerio de Ciencia, Innovaci\'on y Universidades (MCIU/AEI) and the European Regional Development Fund (ERDF) with reference PID-2022-136653NA-I00 (DOI:10.13039/501100011033). DJ also acknowledges support from the Agencia Estatal de Investigaci\'on del Ministerio de Ciencia, Innovaci\'on y Universidades (MCIU/AEI) and the the European Union NextGenerationEU/PRTR with reference CNS2023-143910 (DOI:10.13039/501100011033). C.M.P. gratefully acknowledge the support of the  Swedish National Space Agency (DNR 65/19, 177/19). J.Š. would like to acknowledge the support from GACR grant 23-06384O. Sa.M. ~acknowledges support from the Spanish Ministry of Science and Innovation
(MICINN) through AEI under the Severo Ochoa Centres of Excellence Programme
2020--2023 (CEX2019-000920-S). PGB acknowledges support by the Spanish Ministry of Science and Innovation with the \textit{Ram{\'o}n\,y\,Cajal} fellowship number RYC-2021-033137-I and the number MRR4032204. HJD acknowledges support from the Spanish Research Agency of the Ministry of Science and Innovation (AEI-MICINN) under grant PID2019-107061GB-C66.

\newline

This paper includes data collected by the TESS mission. Funding for the {\sc TESS} mission is provided by the NASA Explorer Program.
The following internet-based resources were used in research for this paper: the SIMBAD database and the VizieR service operated at CDS, Strasbourg, France; the NASA's Astrophysics Data System Bibliographic Services.
This work has made use of data from  the European Space Agency (ESA) mission {\it Gaia} (\url{https://www.cosmos.esa.int/gaia}), processed by the {\it Gaia} Data Processing and Analysis Consortium (DPAC, \url{https://www.cosmos.esa.int/web/gaia/dpac/consortium}). Funding for the DPAC has been provided by national institutions, in particular the institutions participating in the {\it Gaia} Multilateral Agreement. This research has made use of the Exoplanet Follow-up Observation Program (ExoFOP; DOI: 10.26134/ExoFOP5) website, which is operated by the California Institute of Technology, under contract with the National Aeronautics and Space Administration under the Exoplanet Exploration Program. This research has made use of the NASA Exoplanet Archive, which is operated by the California Institute of Technology, under contract with the National Aeronautics and Space Administration under the Exoplanet Exploration Program.
Based on observations made with the Nordic Optical Telescope (NOT), owned in collaboration by the University of Turku and Aarhus University, and operated jointly by Aarhus University, the University of Turku and the University of Oslo, representing Denmark, Finland and Norway, the University of Iceland and Stockholm University at the Observatorio del Roque de los Muchachos, La Palma, Spain, of the Instituto de Astrofisica de Canarias.  We are extremely grateful to the NOT staff members for their unique and precious support during the observations.
This project has received funding from the European Union's Horizon 2020 research and innovation programme under grant agreement No 730890 (OPTICON). This material reflects only the authors views and the Commission is not liable for any use that may be made of the information contained therein.
This work is done under the framework of the KESPRINT collaboration
(http://www.kesprint.science). KESPRINT is an international consortium
devoted to the characterization and research of exoplanets discovered with
space-based missions.

\end{acknowledgements}

\bibliographystyle{aa}
\bibliography{lmbinary}

\appendix
\onecolumn
\section{SED models}
\begin{figure*}[h!]
    \centering
    \includegraphics[width=\linewidth]{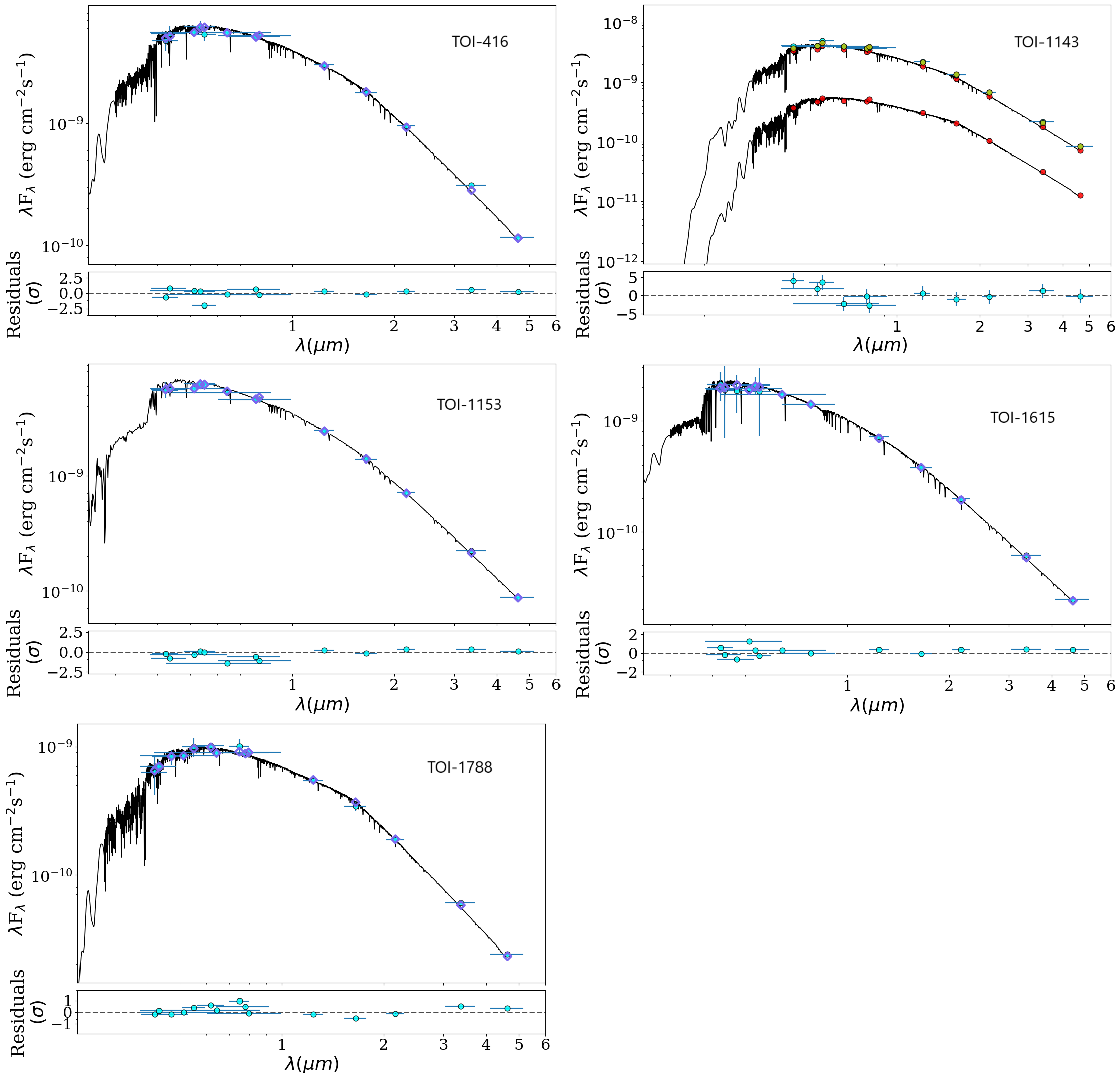}
    \caption{SED fits of primary stars in our systems after subtracting the fluxes of the secondaries. In the case of TOI-1143, we simultaneously modeled both primary stars and fainter companions visible in the speckle imaging, assuming the same distance and contrast $\Delta \mathrm{I}=2.07(10)$.}
    \label{fig:SED}
\end{figure*}

\section{Measured radial velocities}
\renewcommand{\arraystretch}{1.05}
\newcolumntype{M}{>{\centering\arraybackslash\footnotesize}c}
\setlength{\tabcolsep}{4pt}
\begin{table*}[b]
\caption{Relative radial velocities of the target stars.}             
\label{Tab:RVs}      
\centering  
\begin{tabular}{MMMMMMMMM}
\hline
BJD     &       RV (km/s)       &       Err (km/s)      &       BJD     &       RV (km/s)  &       Err (km/s)      &       BJD     &       RV (km/s)       &       Err (km/s)  \\ \hline \hline
\multicolumn{3}{M}{TOI-416 (OES)}                                       &       \multicolumn{3}{M}{TOI-1143 (OES)}                                  &       \multicolumn{3}{M}{TOI-1615 (OES)}                                  \\
2458523.2280    &       19.623  &       0.123   &       2459044.4553    &       -0.062  &       0.085   &       2458892.3214    &       73.955  &       0.119   \\
2458523.2598    &       21.790  &       0.129   &       2459044.5457    &       0.568   &       0.101   &       2458892.5952    &       68.857  &       0.172   \\
2458529.2246    &       12.212  &       0.084   &       2459062.3901    &       12.655  &       0.087   &       2458895.2921    &       24.748  &       0.559   \\
2458530.2746    &       20.736  &       0.183   &       2459068.5359    &       9.758   &       0.092   &       2458898.4056    &       -5.719  &       0.221   \\
2458530.2959    &       20.189  &       0.183   &       2459070.5436    &       16.930  &       0.117   &       2458898.4425    &       -4.858  &       0.124   \\
2458531.2708    &       18.531  &       0.153   &       2459071.5622    &       16.471  &       0.115   &       2458901.3592    &       19.702  &       0.140   \\
2458531.3025    &       16.264  &       0.117   &       2459075.5707    &       8.001   &       0.205   &       2458924.3690    &       58.159  &       0.104   \\
2458532.2923    &       8.149   &       0.116   &       2459089.5950    &       0.287   &       0.128   &       2458925.3568    &       0.000   &       0.095   \\
2458532.3241    &       7.351   &       0.157   &       2459104.5956    &       16.941  &       0.080   &       2458926.4010    &       10.002  &       0.100   \\
2458533.2525    &       0.000   &       0.048   &       2459105.5086    &       15.585  &       0.099   &       2458928.3504    &       55.504  &       0.099   \\
2458533.2947    &       0.240   &       0.052   &       2459106.5041    &       14.089  &       0.077   &       2458930.6063    &       26.474  &       0.353   \\
2459246.2463    &       17.656  &       0.222   &       2459108.4451    &       8.961   &       0.079   &       2458935.5544    &       73.286  &       0.111   \\
\multicolumn{3}{M}{TOI-416 (FIES)}                                      &       2459120.5219    &       6.360   &       0.113   &       2458936.5892    &       33.280  &       0.105   \\
2458513.3854    &       -0.011  &       0.011   &       2459175.6687    &       9.199   &       0.109   &       2458937.5765    &       -8.927  &       0.100   \\
2458514.3988    &       6.396   &       0.010   &       \multicolumn{3}{M}{TOI-1143 (TRES)}                                 &       2458946.6006    &       39.485  &       0.101   \\
2458515.4654    &       16.773  &       0.018   &       2458820.0399    &       -14.525 &       0.063   &       2458947.5723    &       71.317  &       0.159   \\
2458517.3940    &       19.634  &       0.014   &       2458837.0365    &       1.096   &       0.058   &       2459242.2406    &       48.322  &       0.168   \\
2458518.4782    &       9.815   &       0.012   &       2458848.0285    &       1.514   &       0.058   &       \multicolumn{3}{M}{TOI-1615 (TLS)}                                  \\
2458522.4553    &       16.610  &       0.010   &       2458853.9270    &       -14.484 &       0.075   &       2458855.4712    &       8.881   &       0.245   \\
2458523.4578    &       22.160  &       0.012   &       2458859.9602    &       0.000   &       0.058   &       2458860.6821    &       71.814  &       0.240   \\
\multicolumn{3}{M}{TOI-416 (SONG)}                                      &       2458871.9689    &       -1.904  &       0.028   &       2458864.3191    &       64.048  &       0.234   \\
2458512.3326    &       23.078  &       0.005   &       2458876.0529    &       -14.264 &       0.081   &       2458865.3602    &       50.574  &       0.244   \\
2458513.3293    &       20.786  &       0.004   &       2458878.8815    &       0.721   &       0.042   &       \multicolumn{3}{M}{TOI-1788 (OES)}                                  \\
2458514.3292    &       26.586  &       0.006   &       2458885.9672    &       -10.306 &       0.071   &       2458959.4386    &       36.293  &       0.177   \\
2458515.3316    &       36.470  &       0.004   &       2458888.9593    &       -6.892  &       0.052   &       2458960.4531    &       20.843  &       0.117   \\
2458516.3566    &       42.637  &       0.005   &       \multicolumn{3}{M}{TOI-1153 (TLS)}                                  &       2458961.3928    &       -0.261  &       0.168   \\
2458517.3148    &       40.522  &       0.004   &       2458923.5398    &       -60.720 &       0.250   &       2458963.4300    &       18.166  &       0.111   \\
2458518.3412    &       32.327  &       0.005   &       2458923.5505    &       -60.010 &       0.380   &       2458964.5088    &       36.929  &       0.206   \\
\multicolumn{3}{M}{TOI-1143 (OES)}                                      &       2458923.5613    &       -59.980 &       0.180   &       2458967.5505    &       1.288   &       0.111   \\
2458892.4386    &       16.322  &       0.086   &       2459067.5052    &       -83.650 &       0.350   &       2458975.4397    &       37.716  &       0.299   \\
2458892.5283    &       16.234  &       0.083   &       2459068.5750    &       -53.510 &       0.190   &       2459191.6999    &       0.000   &       0.098   \\
2458895.3833    &       9.790   &       0.053   &       2459094.4663    &       5.190   &       0.230   &       2459246.4524    &       14.752  &       0.463   \\
2458901.3009    &       14.855  &       0.072   &       2459095.5200    &       -16.310 &       0.240   &       2459983.4748    &       8.991   &       0.095   \\
2458924.4007    &       16.866  &       0.083   &       2459099.5398    &       -18.110 &       0.370   &       2460004.4211    &       3.283   &       0.147   \\
2458925.3960    &       16.375  &       0.056   &       2459100.4393    &       4.750   &       0.230   &       2460065.3497    &       38.356  &       0.143   \\
2458930.3286    &       5.452   &       0.108   &       2459100.6216    &       5.110   &       0.270   &       2460087.3911    &       32.868  &       0.399   \\
2458930.4164    &       5.111   &       0.113   &       2459101.6213    &       -19.350 &       0.270   &       \multicolumn{3}{M}{TOI-1788 (TLS)}                                  \\
2458931.3062    &       1.785   &       0.059   &       2459177.5857    &       -37.320 &       0.280   &       2459177.6200    &       25.944  &       0.364   \\
2458931.3936    &       1.609   &       0.073   &       2459177.5967    &       -36.440 &       0.560   &       2459177.6417    &       26.178  &       0.257   \\
2458932.3371    &       -0.178  &       0.051   &       2459178.6741    &       2.030   &       0.510   &       2459265.5971    &       4.600   &       0.175   \\
2458932.4288    &       0.000   &       0.046   &       2459179.4647    &       0.780   &       0.740   &       2459265.6188    &       4.152   &       0.135   \\
2458933.3227    &       3.349   &       0.057   &       2459179.4758    &       0.790   &       0.660   &       2459268.5539    &       27.443  &       0.161   \\
2458933.4101    &       4.057   &       0.056   &       2459212.5910    &       -81.020 &       0.310   &       2459270.4744    &       14.425  &       0.114   \\
2458936.3252    &       16.814  &       0.061   &       2459214.4187    &       -11.000 &       0.250   &       2459271.4616    &       -5.222  &       0.238   \\
2458936.3674    &       16.789  &       0.063   &       2459235.4186    &       -64.210 &       0.330   &       2459277.3885    &       -3.663  &       0.095   \\
2458940.6134    &       7.463   &       0.080   &       2459246.5492    &       -21.920 &       0.230   &       2459299.5060    &       5.240   &       0.112   \\
2458956.4440    &       8.543   &       0.061   &       2459266.6406    &       -83.720 &       0.270   &       2459303.5632    &       -3.143  &       0.098   \\
2458957.4578    &       15.449  &       0.071   &       2459266.6517    &       -83.450 &       0.220   &       2459304.5609    &       0.552   &       0.086   \\
2458991.3658    &       16.101  &       0.091   &       2459267.5547    &       -61.740 &       1.310   &       2459304.5827    &       1.008   &       0.051   \\
2459002.4285    &       15.758  &       0.076   &       2459267.5658    &       -61.290 &       1.040   &       2459305.5689    &       19.933  &       0.142   \\
2459016.5113    &       14.793  &       0.121   &       2459268.5710    &       -16.850 &       0.120   &       2459305.5906    &       20.017  &       0.217   \\
2459035.3838    &       12.755  &       0.102   &       2459269.6577    &       5.100   &       0.190   &       2459309.5167    &       -3.738  &       0.189   \\
2459040.4020    &       11.440  &       0.090   &       2459270.5375    &       -15.470 &       0.300   &               &               &               \\
2459040.4890    &       11.188  &       0.081   &       2459271.5028    &       -59.130 &       0.350   &               &               &               \\
2459040.5312    &       10.929  &       0.079   &       2459277.3644    &       -51.350 &       0.200   &               &               &               \\ \hline
                                                                
\end{tabular}
\end{table*}

\end{document}